\documentclass[5p,times,twocolumn]{cas-dc}

\usepackage[numbers]{natbib}
\definecolor{mygray}{gray}{.9}
\definecolor{myblue}{RGB}{93,80,180}
\definecolor{mygreen}{RGB}{93,173,85}

\def\tsc#1{\csdef{#1}{\textsc{\lowercase{#1}}\xspace}}
\tsc{WGM}
\tsc{QE}
\begin{document}
\let\WriteBookmarks\relax
\def\floatpagepagefraction{1}
\def\textpagefraction{.001}

\shorttitle{A Novel Scene Coupling Semantic Mask Network for Remote Sensing Image Segmentation}    

\shortauthors{X.Ma et al.}  

\title [mode = title]{A Novel Scene Coupling Semantic Mask Network for Remote Sensing Image Segmentation}

\author[1,2]{Xiaowen Ma}[type=editor,orcid=0000-0001-5031-2641]%[<options>]
\ead{xwma@zju.edu.cn}

\ead[url]{https://scholar.google.com.hk/citations?user=UXj8Q6kAAAAJ&hl=zh-CN}
\author[1]{Rongrong Lian}
\ead{lianrr@zju.edu.cn}
\author[1]{Zhenkai Wu}
\ead{zkwu@zju.edu.cn}
\author[6]{Renxiang Guan}
\ead{renxiangguan@nudt.edu.cn}
\author[1]{Tingfeng Hong}
\ead{tfhong@zju.edu.cn}
\author[1]{Mengjiao Zhao}
\ead{22351275@zju.edu.cn}
\author[3]{Mengting Ma}
\ead{mtma@zju.edu.cn}
\author[4]{Jiangtao Nie}
\ead{niejiangtao@mail.nwpu.edu.cn}
\author[5]{Zhenhong Du}
\ead{duzhenhong@zju.edu.cn}
\author[7]{Siyang Song}
\ead{ss2796@cam.ac.uk}
\cormark[1]
\author[1,8]{Wei Zhang}
\ead{cstzhangwei@zju.edu.cn}
\cormark[1]
\cortext[1]{Corresponding author}



\affiliation[1]{organization={School of Software Technology, Zhejiang University},
            addressline={}, 
            city={Hangzhou},
%          citysep={}, % Uncomment if no comma needed between city and postcode
            postcode={310027}, 
            country={China}}
\affiliation[2]{organization={Noah's Ark Lab, Huawei},
            addressline={}, 
            city={Shanghai},
%          citysep={}, % Uncomment if no comma needed between city and postcode
            postcode={201206}, 
            country={China}}
\affiliation[3]{organization={School of Computer Science and Technology, Zhejiang University},
            addressline={}, 
            city={Hangzhou},
%          citysep={}, % Uncomment if no comma needed between city and postcode
            postcode={310027}, 
            country={China}}
\affiliation[4]{organization={School of Computer Science and Technology, Northwestern Polytechnical University},
            addressline={}, 
            city={Xian},
%          citysep={}, % Uncomment if no comma needed between city and postcode
            postcode={710072}, 
            country={China}}
\affiliation[5]{organization={School of Earth Sciences, Zhejiang University},
            addressline={}, 
            city={Hangzhou},
%          citysep={}, % Uncomment if no comma needed between city and postcode
            postcode={310027}, 
            country={China}}
\affiliation[6]{organization={School of Computer Science and Technology, National University of Defense Technology},city={Changsha},postcode={410073},city={China}}
\affiliation[7]{organization={School of Computing and Mathematical Sciences, University of Leicester},
            addressline={}, 
            % city={Hangzhou},
%          citysep={}, % Uncomment if no comma needed between city and postcode
            % postcode={310027}, 
            country={UK}}
% \author[2]{}%[
% % Address/affiliation
\affiliation[8]{organization={Innovation Center of Yangtze River Delta, Zhejiang University},
            addressline={}, 
            city={Jiaxing},
%          citysep={}, % Uncomment if no comma needed between city and postcode
            postcode={314103}, 
            country={China}}
% % Corresponding author text

% % Footnote text


\begin{abstract}
As a common method in the field of computer vision, spatial attention mechanism has been widely used in semantic segmentation of remote sensing images due to its outstanding long-range dependency modeling capability. However, remote sensing images are usually characterized by complex backgrounds and large intra-class variance that would degrade their analysis performance. While vanilla spatial attention mechanisms are based on dense affine operations, they tend to introduce a large amount of background contextual information and lack of consideration for intrinsic spatial correlation. To deal with such limitations, this paper proposes a novel scene-Coupling semantic mask network, which reconstructs the vanilla attention with scene coupling and local global semantic masks strategies. Specifically, \textbf{scene coupling} module decomposes scene information into global representations and object distributions, which are then embedded in the attention affinity processes. This Strategy effectively utilizes the intrinsic spatial correlation between features so that improve the process of attention modeling. Meanwhile, \textbf{local global semantic masks} module indirectly correlate pixels with the global semantic masks by using the local semantic mask as an intermediate sensory element, which reduces the background contextual interference and mitigates the effect of intra-class variance. By combining the above two strategies, we propose the model SCSM, which not only can efficiently segment various geospatial objects in complex scenarios, but also possesses inter-clean and elegant mathematical representations. Experimental results on four benchmark datasets demonstrate the the effectiveness of the above two strategies for improving the attention modeling of remote sensing images. 
For example, compared to the recent advanced method LOGCAN++, the proposed SCSM has 1.2\%, 0.8\%, 0.2\%, and 1.9\% improvements on the LoveDA, Vaihingen, Potsdam, and iSAID datasets, respectively. The dataset and code are available at \url{https://github.com/xwmaxwma/rssegmentation}.
\end{abstract}

% Keywords

\maketitle

\section{Introduction}

Semantic segmentation, also known as semantic labeling, is one of the fundamental and challenging tasks in remote sensing image understanding. It aims to assign a pixel-wise semantic class label to the given image, which plays a crucial role in urban planning \cite{urban,urban2,he2024geolocation}, environmental protection \cite{environment,cui2024real}, and natural landscape monitoring~\cite{road, wang2024airshot,wang2024onls}. Meanwhile, land cover information of various geospatial objects usually provides crucial insights from a panoramic perspective to tackle a multitude of socioeconomic and environmental challenges, such as food insecurity, poverty, climate change, and disaster risk. With recent advances in earth observation technologies, a constellation of satellite and airborne platforms have been launched, resulting in substantial fine-resolution remotely sensed images available for semantic segmentation \cite{rsdataset,liu2024crossmatch, 10647298,cui2024superpixel,cui2022unsupervised, chen2024bimcv, chen2024learning}, allowing to better achieve the above applications. To manifest the physical properties of land cover, vegetation indices have been frequently used as the measure for extracting features from multi-spectral/multi-temporal images \cite{significant}, despite that the adaptability and flexibility of these indices are severely restricted by their high dependency on hand-crafted descriptors \cite{fcn}.

\begin{figure*}[t]
	\centering	\includegraphics[width=0.99\textwidth]{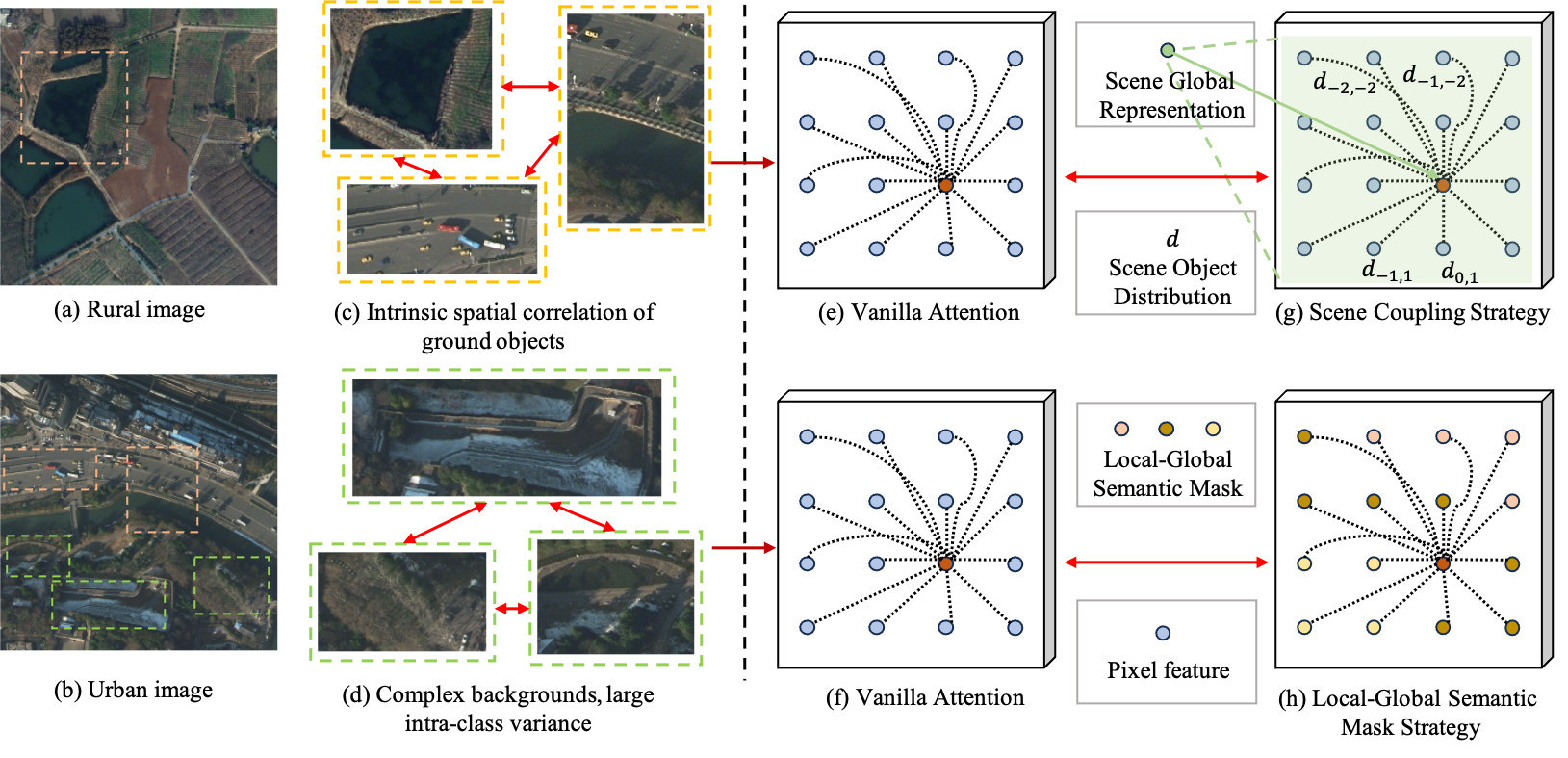}
	\caption{Intuitive understanding of the scene coupling and semantic masks. For remote sensing images recorded in two different scenerios, i.e., (a) rural images and (b) urban images, we first give two examples to represent (c) the intrinsic spatial correlation of remote sensing image feature targets and (d) the complex backgrounds, large intra-class variance, respectively. For the former, we design Scene-Coupling Attention to gain context modeling by embedding scene global representations and scene object distributions in the (e) Vanilla Attention. For the latter, we introduce (h) local-global semantic mask strategy with a spatial prior, which can avoid large background noise interference and mitigate intra-class variance compared with (f) Vanilla Attention.}
	\label{fig:intro}
\end{figure*}

With the recently proposed deep learning models \cite{fcn,transformer,zhang2023body,wang2024sparse,nie2024imputeformer}, plenty of dramatic breakthroughs of semantic segmentation have been made. Compared to hand-crafted methods that merely take finite bands into account \cite{significant,trias2008using}, Convolution Neural Networks (CNNs) can learn more comprehensive task-specific cues from periods, spectrum, and the interactions between different categories of land cover. 
However, due to the fixed geometric structure, CNNs are limited by local receptive fields and short-range contextual information. Therefore, a series of subsequent approaches devoted to context modeling \cite{jin2022imc,bi2024decoding,chen2023self,qianmaskfactory}, including spatial context modeling \cite{pspnet,deeplabv3+} and relational context modeling \cite{ocrnet,chen2024tokenunify}, have been proposed to capture long-range dependencies. Specifically, these spatial context modeling approaches aim to enhance the pixel-level feature representations by employing spatial pyramid pooling~\cite{pspnet} and dilated convolution~\cite{deeplabv3+} to integrate spatial context information. Besides, DMNet~\cite{dmnet} introduces input adaptively generated multi-scale convolution kernels to extract multi-scale features. However, such approaches focus on capturing homogeneous context dependencies and thus tend to ignore categorical differences among different geospatial objects, which may lead to unreliable contexts if there are confusing categories (e.g., road and barren) in the scene \cite{isnet}. 

Alternatively, relational context modeling approaches \cite{logcan++,ocrnet} are mainly built on the attention mechanism, which achieve promising results in semantic segmentation tasks by computing pixel-level similarity in images and weighting aggregation of heterogeneous contextual information. They utilized various strategies to improve the attention modeling process, including spatial self-attention and channel self-attention~\cite{danet,senet}, class attention \cite{ocrnet,docnet}, efficiency improvements~\cite{ccnet}, and fully attention \cite{flanet}. 
For example, DANet \cite{danet} enhances the semantic representation of features by modeling the spatial similarity and channel similarity between pixel features. However, higher computational complexity and large number of irrelevant background contexts limit the performance of the model. 
FLANet \cite{flanet} innovatively proposes full attention, modeling both spatial and pixel dependencies in a single module in order to reduce the attention deficit problem that occurs with single attention \cite{flanet}. Similarly, it is severely interfered by a large amount of background noise.
OCRNet \cite{ocrnet} effectively reduces the interference of background contexts by proposing a global class representation. 
However, considering the large intra-class variance in remote sensing images, the large semantic distance between pixels and the global class representation impairs the context modeling performance. 
Overall, since these approaches usually depend on dense affine operations, a large amount of background contexts are introduced to blur their foreground features, causing impaired recognition of geospatial objects \cite{pointflow}.
In addition, these approaches fail to consider the intrinsic spatial correlation between geospatial objects of remotely sensed images and high intra-class variance in the attentional affinity process, and thus their segmentation performances are limited. Although recently proposed transformer-based solutions \cite{transformer} and state space models \cite{mamba,mamba2} have outstanding ability in modeling global contexts for remote sensing images, these transformers \cite{swinunet,zhang2022transformer,ding2022looking,glots,sun2024ultra,cheng2024spt} generally rely on vanilla attention, and thus still possess the aforementioned shortcomings (i.e., dense affinity and ignoring spatial correlation of geospatial objects). Moreover, the quadratic computational complexity and memory consumption makes the vanilla attention operation involved in them unrealistic to be applied to high-resolution remote sensing images. Meanwhile, state space models \cite{rsmamba} usually not only suffer from low inference speed, but also rely on expensive GPU and selective scanning mechanisms, making them not favorable for practical applications.

In this paper, we propose a novel scene coupling semantic mask network which can accurately segment various geospatial objects in complex scenes in an efficient manner, where an restructured attention operation with novel scene-coupling and semantic mask strategies is proposed as the decoder head. Specifically, we first reconfigure the affinity process of vanilla attention based on the scene coupling strategy, which exploits the intrinsic spatial correlation between geospatial objects to improve the attention modeling process. Scene-coupling attention strives to decompose a scene into a scene-global representation and a scene-object distribution, which are then embed in the attention computation process to model the two concurrent modes exhibited by correlations in remotely sensed images (i.e., rural and urban, which are depicted in Section \ref{sec-finding}). In addition, In addition, we propose a preprocessing method that enhances the vanilla attention mechanism by reconstructing its input using local and global semantic masks. It proposes local semantic masks with spatial prior as intermediate perceptual elements to indirectly correlate pixels and global semantic masks, which avoids the interference of background noise and mitigates intra-class variance.

In particular, by combining the two elegantly, the proposed SCSM model not only achieves state-of-the-art performance on four remote sensing benchmark segmentation datasets, but also possesses a concise formulaic representation. The main contributions of this paper are summarized as follows.
\begin{itemize}

\item We propose a novel scene coupling attention mechanism, which benefits attention modeling due to the introduce of intrinsic spatial correlation between the ground-object targets. In particular, a novel image-level rotary position encoding module ROPE+ is devised to model object distribution within a scene. It allows the model to understand the absolute positions of targets and their relative distances very simply.

\item We explore the scene global representation from the frequency domain, which is the first dual-domain attention model for semantic segmentation of remote sensing images.

\item We introduce a local global semantic mask strategy with spatial prior by considering the feature of large intra-class variance of complex background of remote sensing images. 

\item Based on the operation of spatial localization, we effectively combine scene coupling attention and semantic mask and propose SCSM. the results on four benchmark datasets show that the proposed SCSM outperforms other state-of-the-art methods and has a better balance between accuracy and efficiency with smaller number of parameters and computational effort.
\end{itemize}

\begin{figure*}[t]
	\centering	\includegraphics[width=0.9\textwidth]{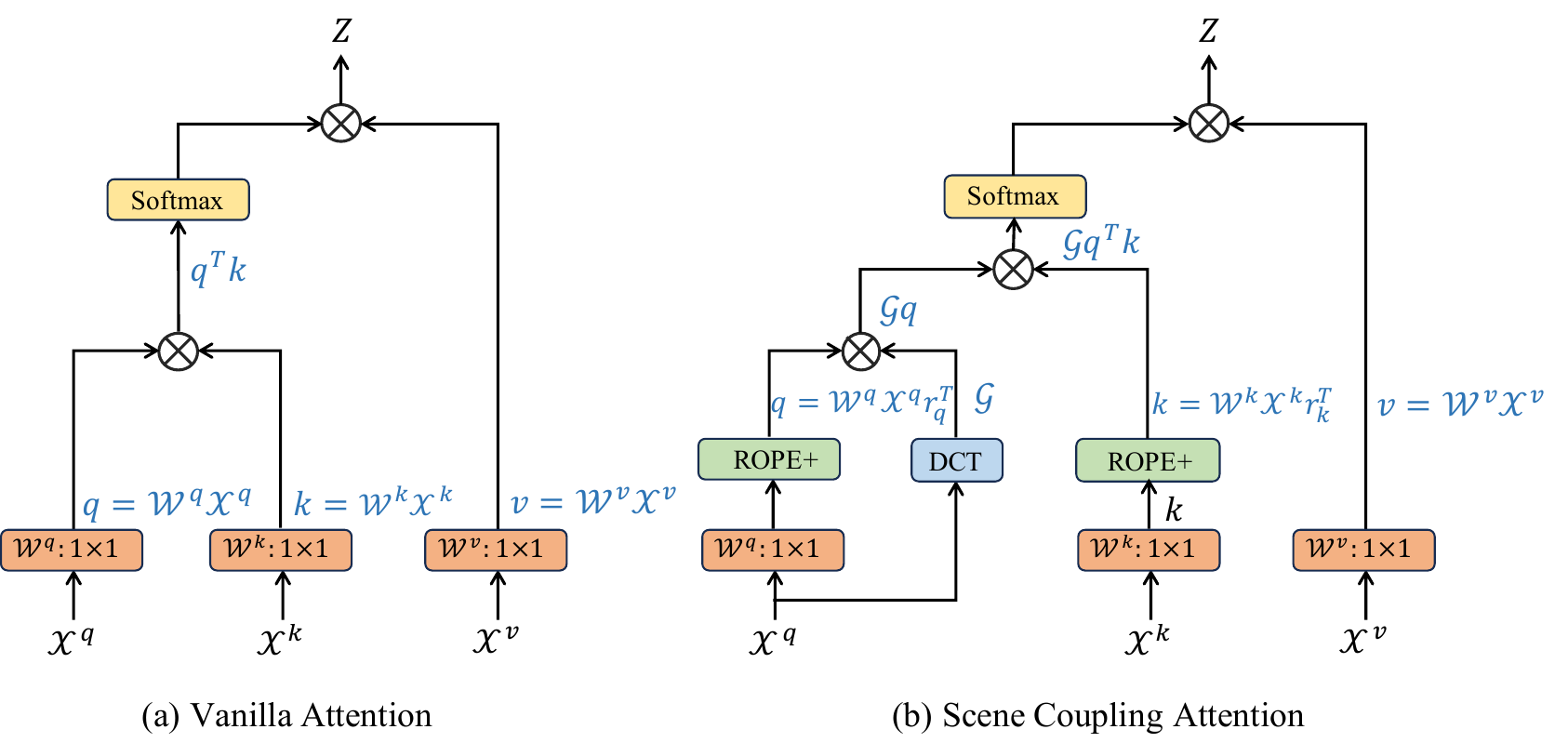}
	\caption{Our proposed Scene Coupling Attention (SCA) enhances the vanilla attention mechanism by incorporating additional positional and global scene representations. It first applies 2D Rotary Position Embedding (ROPE+) to both query and key, indirectly modeling the relative spatial distribution of objects within the scene. Additionally, it applies a 2D Discrete Cosine Transform (DCT) to the query to obtain a global scene representation, which is then integrated to effectively capture the intrinsic spatial correlations between objects.}
	\label{attn}
\end{figure*}

\section{Related Work}

\subsection{General Semantic Segmentation}

Methods based on fully convolutional networks (FCNs) have made great progress in semantic segmentation by exploiting the powerful representation capabilities of classification networks~\cite{resnet} pre-trained on large-scale data. Subsequent work has been devoted to contextual modeling and thus enhancing feature discrimination. For example, PSPNet~\cite{pspnet} generates pyramidal feature maps by pyramid pooling, Deeplab~\cite{deeplabv3+} proposes dilated convolution kernels that force the network to perceive larger regions, and DenseASPP~\cite{denseaspp} further increases the expansion rate based on Deeplab. In addition, some works acquire the global context of the input image based on the attention mechanism~\cite{danet,ccnet,ocrnet,flanet,caa}. 
For example, DANet \cite{danet} introduces parallel spatial attention and channel attention, CCNet \cite{ccnet} proposes criss-cross attention to improve computational efficiency. OCRNet \cite{ocrnet} and CAANet \cite{ccanet} propose to use class centers to participate in attention computation. 
Also, some other works such as GMMSeg~\cite{gmmseg} and Protoseg~\cite{protoseg} propose to generate prototypes based on Gaussian mixture models or online averaging to improve the segmentation performance.
In recent years, some transformer methods have gained popularity, and ViT~\cite{vit} has obtained remarkable results by migrating transformer to the image recognition field for the first time. 
In order to optimize image recognition tasks such as segmentation, some of the next works improved this, such as hierarchical structure~\cite{segformer,swintransformer}, token mixer improvement~\cite{mlp,mlps}, decoder design~\cite{maskformer,wu2024domain}, training strategy optimization \cite{beit,sunprogram} and complexity reduction~\cite{dilateformer}. 
For example, Swin Transformer \cite{swintransformer} introduces a windowing mechanism and multi-scale design to extract local context and reduce the computational complexity of vanilla attention. Segmenter \cite{segmenter} proposes a transformer-based decoder to generate class masks with good scalability to perform a wider range of semantic segmentation tasks. SegFormer \cite{segformer} combines a hierarchical transformer with a lightweight multilayer perceptron that avoids the limitations of positional coding on the resolution of the input image and presents a robust representation. 
However, these attention-based general semantic segmentation methods \cite{danet,ocrnet,swintransformer} ignore the large intra-class variance characteristic of correlation and background complexity among remote sensing image feature targets, and thus achieve unsatisfactory results.

\subsection{Semantic Segmentation in Remote Sensing Community}

The remote sensing community has many applications for semantic segmentation, such as road extraction~\cite{road1,bdtnet,bmda}, building detection~\cite{building1, lbe}, land use and land cover classification~\cite{land2,jdp}. 
Specially, RADANet \cite{road1} develops a deformable attention-based network to learn the remote dependencies of specific road pixels. BEM \cite{building1}  introduces overall nested edge detection to extract edge features thereby enhancing the building boundary extraction capability.
These methods follow the general semantic segmentation for specific application scenarios (e.g., roads or buildings), while improving them. Due to the specific segmentation objects, these methods cannot be better generalized to other application scenarios. Moreover, for non-specific application scenarios, some attention-based methods~\cite{mdanet,SSGCC,CMSCGC,manet,ccanet} have obtained superior performance, such as deformable attention~\cite{mdanet}, linear attention~\cite{manet}, class attention~\cite{ccanet}, etc. These works have to some extent improved the spatial attention mechanism for application to the remote sensing community. However, these works do not consider the large intra-class variance of complex backgrounds and ignore the correlation between feature targets due to the lack of analysis of the characteristic differences between natural images and remotely sensed images. Some recent works~\cite{farseg, pointflow,sco} have this difference analysis and designed model structures purposefully, then they are mainly based on foreground aware~\cite{farseg}, contrastive learning \cite{sco} or sparse mapping~\cite{pointflow}, which is different from the starting point of this paper to improve the attention mechanism in remote sensing communities. Therefore, this paper proposes a general semantic segmentation approach for the remote sensing community, i.e., a scene Coupling semantic mask network, to improve the performance of spatial attention mechanisms in the field of remote sensing image segmentation.

\section{Preliminaries and Findings}

In this section, we first depict the theoretical computational process of the vanilla attention \cite{danet,nonlocal} to help the understanding of the subsequent optimization mechanism of scene-coupled semantic masking strategy. Subsequently, we systematically analyze the properties of remotely sensed images, based on which the corresponding findings in conjunction with the theoretical descriptions in the Preliminaries (Eq.1-Eq.5) is obtained, i.e., the vanilla attention mechanism usually performs unsatisfactorily due to dense affinity, the neglect of intra-class variance and intrinsic spatial correlation among geospatial objects for segmentation reasons. 
%This motivates us to propose SCSM, which reconstruct vanilla attention through scene coupling and semantic masking strategies thereby enhancing the performance of the attention model on remote sensing image segmentation tasks.

\subsection{Preliminaries}

%To facilitate the subsequent analysis of the inadequacy of vanilla attention \cite{nonlocal,danet} module (VAM) for remote sensing image segmentation tasks, we first depict the process of its theoretical computation.

As shown in Fig. \ref{attn}, given a feature map $\boldsymbol{x} \in \mathbb{R}^{C \times N}$, where $N = H \times W$, $C$ representing the number of channels while $H$ and $W$ indicate its height and width, respectively, the initial processing of vanilla attention \cite{nonlocal,danet} module (VAM) involves the application of a projection matrix, where query, key, and value can be obtained as:
\begin{equation}
\label{eq-begin-1}
\begin{split}
    \boldsymbol{q} &= f_q(\boldsymbol{x^q})=\boldsymbol{W}_q\boldsymbol{x}^q, \\
    \boldsymbol{k} &= f_k(\boldsymbol{x^k})=\boldsymbol{W}_k\boldsymbol{x}^k,\\
    \boldsymbol{v} &= f_v(\boldsymbol{x^v})=\boldsymbol{W}_v\boldsymbol{x}^v,      
\end{split}
\end{equation}
where $\boldsymbol{W}_q$, $\boldsymbol{W}_k$, and $\boldsymbol{W}_v$ represent the corresponding projection matrices for encoding query, key and value respectively, which are commonly used to normalize the channel number of feature maps to facilitate attention computation.

For ease of subsequent description, this paper assumes that these projection matrices are of size $C \times C$, maintaining the number of channels consistent for pre and post-projection. In vanilla spatial attention operators, $\boldsymbol{x}^q$, $\boldsymbol{x}^k$, and $\boldsymbol{x}^v$ are identical to $\boldsymbol{x}$ as:
\begin{equation}
\label{eq-begin-2}
    \boldsymbol{x}^q=\boldsymbol{x},\quad \boldsymbol{x}^k=\boldsymbol{x},\quad \boldsymbol{x}^v=\boldsymbol{x}.
\end{equation}
Subsequently, the module calculates the similarity between queries and keys, and performs dot-product scaling, which can be formulated as:
\begin{equation}
\label{eq-begin-3}
   t_{m,n}=\frac{{\boldsymbol{q}^\intercal_m} \boldsymbol{k}_n}{\sqrt{C}} = \frac{{f_q(\boldsymbol{x})}_m^\intercal{f_k(\boldsymbol{x})}_n}{\sqrt{C}},
\end{equation}
where $t \in \mathbb{R}^{N \times N}$ is the similarity matrix between query and key, and a normalization is then applied along the last dimension using the Softmax function:
\begin{equation}
   \alpha_{m,n}=\frac{{e}^{t_{m,n}}}{\sum_{\hat{n}=1}^{N}{e}^{t_{m,\hat{n}}}}.
\end{equation}
This way, attention weights are normalized to drive the model to focus on the features of important regions.

Finally, the VAM computes the output $Z$ by weighting the values based on the normalized similarity matrix as:
\begin{equation}
    {Z}_{i}=\sum_{j=1}^{N}\alpha_{ij}\boldsymbol{v}_j.
\end{equation}
With the above operations, the VAM can selectively capture different global contexts based on the similarity between all pixels within an image, thereby enhancing the feature representation of the pixels.

Despite the remarkable achievements of VAM on natural images \cite{vit,danet}, the performance on remote sensing images is usually unsatisfactory \cite{pointflow}. As analyzed in the introduction, remote sensing images exhibit three distinct characteristics: complex backgrounds, large intra-class variance, and inherent spatial correlations among ground objects. VAM is operated based on dense affinity calculations (i.e., Eq.2), assess similarities and aggregate them across image pixels, inadvertently incorporating a substantial amount of background context. Additionally, the discrete computation of similarity does not account for the interrelations among ground object targets, leading to suboptimal segmentation performance. Therefore, addressing these three characteristics, this paper proposes scene Coupling and local-global semantic mask reconstructed spatial attention to enhance model performance.

\begin{figure*}[t]
	\centering	\includegraphics[width=0.8\textwidth]{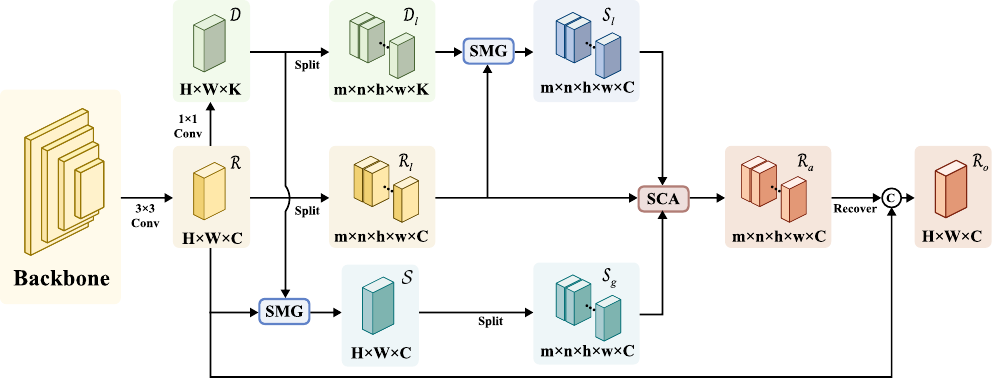}
	\caption{Overall structure diagram of the SCSM model, which consists of backbone, several convolution operations, two semantic mask generation (SMG) modules and a scene coupling attentio (SCA) module. The SMG module associates pixels with the global semantic mask through a local mask spatial prior, achieving class-level modeling and seamless integration with the scene coupling module. The SCA module generates a global scene representation and object distribution, embedding them into the attention calculation to effectively uncover correlations between feature targets in remote sensing images. Overall, the workflow of SCSM is as follows: 1, the backbone processes the input image to extract the deep feature $\mathcal{R}$, which has a resolution of 1/8 of the original image; 2, the deep feature $\mathcal{R}$ is input to the SMG and SCM modules to obtain the class-wise context enhanced feature $\mathcal{R}_o$; and 3, $\mathcal{R}_o$ is processed with $1 \times 1$ convolution to obtain the classification representation, followed by a bilinear interpolation based up-sampling to obtain the final segmentation mask.}
	\label{scsm}
\end{figure*}

\subsection{Findings}
\label{sec-finding}
In this subsection, we further provide a typical visual example in Fig. \ref{fig:intro}, where two common scenarios in remote sensing images, i.e., rural and urban, are analyzed. Note that more visual cases are widely available in various remote sensing datasets \cite{vaihingen,loveda,isaid}. 

We decompose the scene into the scene global representation and the scene object distribution, which are the basic elements that make up a scene. By integrating these two elements through deep learning techniques \cite{transformer,resnet}, we are able to model different scenes in the real world. To facilitate the understanding of our statement, we give a typical visual example as shown in Fig. \ref{fig:intro}. 
Firstly, remote sensing image segmentation mainly targets to geospatial objects that usually have strong dependencies with their scenes. As illustrated in Fig. \ref{fig:intro}(c), the examples of scene dependence can be divided into two categories: (1) For the interior of the scene, objects in close proximity usually show some combination or concurrence, and pixels near the objects may show a high degree of correlation. For example, cars tend to be parked on the road, while buildings are usually distributed on both sides of the road; and (2) For different scenes, the pairwise relationships between pixel features may different. For example, in rural scenes water is surrounded by farmland, while in urban scenes, water is generally distributed next to roads.

However, as shown in Fig. \ref{fig:intro} (e), VAM is calculated based on independent correlations and does not consider scenes. In other words, VAM ignores such correlations between geospatial objects as depicted above during the attentional computation process, which impairs the performance of attentional modeling. Therefore, we propose scene Coupling attention to address this shortcoming., as in Fig. \ref{fig:intro}(g). 

Specifically, we decompose the scenes into two parts, the scene object distribution as well as the scene global information. 
Firstly, we mine the rich spectral information of the scene based on the discrete cosine transform to extract the scene global representation. Then, we propose the Image-level rotational position encoding (ROPE+) module to model the scene object distribution, and integrate it directly into the attention calculation process. By decomposing scenes into scene-global representations and scene-object distributions, combined with an attention mechanism integrated into a unified module, we gain contextual modeling by effectively exploiting the intrinsic spatial correlation between the ground-object targets.

In addition, as shown in (d) in Fig. \ref{fig:intro}, remote sensing images frequently suffer from complex backgrounds and large intra-class variance characteristics \cite{farseg,farseg++,pointflow} (despite belonging to both background classes, yet they have significant spectral and texture differences). The dense affinity operation-based vanilla spatial attention mechanism usually tends to introduce a large amount of background context, which makes the segmentation performance unsatisfactory \cite{pointflow}. To this end, we propose a novel local global semantic mask with spatial prior, which uses representative semantic masks as the key and value of attention, and indirectly associates pixels with the global semantic mask by using the local semantic mask as an intermediate perceptual element. This allows the proposed model to effectively avoid the interference of background context and mitigates the intra-class variance in the context modeling process.

\begin{figure*}[t]
    \centering \includegraphics[width=0.95\textwidth]{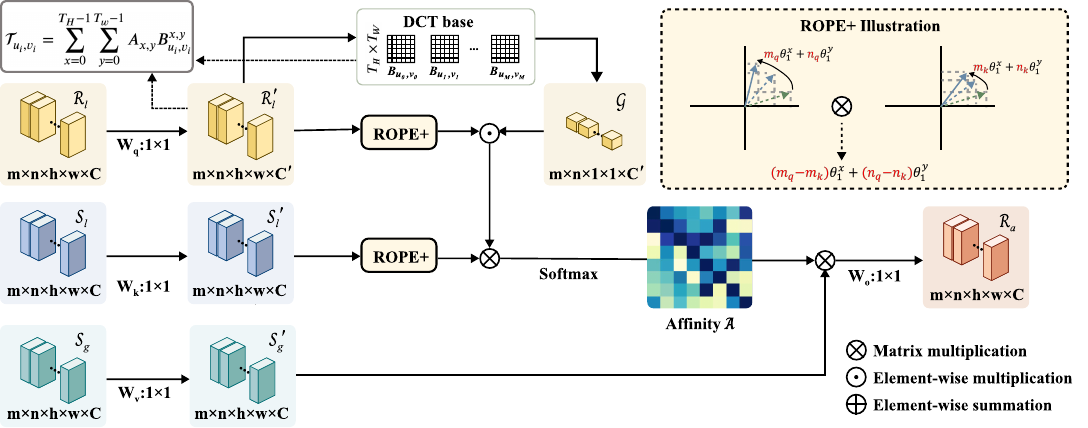}
    \caption{Structural diagram of the Scene Coupling Attention Module, composed of scene object distribution and global scene representation. The scene object distribution leverages ROPE+ positional encoding to capture complex object distributions in remote sensing scenes. The global scene representation is achieved by converting spatial information to frequency domain using 2D DCT, followed by channel attention enhancement to capture valuable spectral information.}
    \label{scm}
\end{figure*}

\section{Methodology}

In this section, we present our Scene Coupling Semantic Mask Network (SCSM) which reconstructs the vanilla attention with scene coupling and local global semantic masks strategies, thus better targeting the characteristics of remote sensing images to enhance segmentation performance..

\subsection{Overall Architecture}

\textbf{Overview:} As depicted in Figure \ref{scsm}, the proposed SCSM is built on an encoder-decoder architecture, which consists of three main components: a backbone, a Semantic Mask Generation (SMG) module, and a Scene Coupling Attention (SCA) module. The SMG module constructs representative local-global semantic masks from features extracted by the backbone, which alleviates the interference of background contexts and issues related to large intra-class variance. Then, we process the pixel features based on the pre-trained DCT base and ROPE+ module to obtain the scene global representations and scene object distributions, respectively. Next, the SCA module embeds scene global representations and scene object distributions into the attention computation process, aiming to effectively extract the correlations among geospatial objects in remote sensing images. By elegantly combining our proposed SMG and SCA modules, vanilla attention (i.e., Eq. \ref{eq-begin-1} and Eq. \ref{eq-begin-2}) is reconstructed with scene coupling (Eq. \ref{eq-final-2}) and semantic masking strategies (Eq. \ref{eq-final-3}). This way, effective modeling of geospatial objects in complex scenes is achieved.

Specifically, the input image $\mathcal{I}$ is firstly fed into the backbone network, and the obtained features are reduced in dimensionality through a $3 \times 3$ convolution to produce the representation $\mathcal{R}$ as: 
\begin{equation}
    \mathcal{R} = \mathrm{Conv}_{3\times3}(\mathrm{Backbone}(\mathcal{I})).
\end{equation}
Then, the $\mathcal{R}$ is transformed to a pre-classification representation $\mathcal{D}$ through a $1 \times 1$ convolution layer in order to get the class distribution of the pixels. Here, $\mathcal{R}$ and $\mathcal{D}$ are further spatially split into $\mathcal{R}_l$ and $\mathcal{D}_l$  respectively to obtain the local semantic mask. Subsequently, $\mathcal{R}$ and $\mathcal{D}$, as well as $\mathcal{R}_l$ and $\mathcal{D}_l$, are fed to the SMG module to generate spatially priorized global semantic masks $\mathcal{S}_g$ and local semantic masks $\mathcal{S}_l$. 
\begin{equation}
    \mathcal{S}_l = \mathrm{SMG}(\mathcal{R}_l,\mathcal{D}_l),\quad \mathcal{S}_g = \mathrm{Split}(\mathrm{SMG}(\mathcal{R},\mathcal{D})),
\end{equation}
where the local class representation $\mathcal{S}_l$, the global class representation $\mathcal{S}_g \in \mathbb{R}^{C \times H \times W}$

Then, $\mathcal{R}_l$, $\mathcal{S}_l$, and $\mathcal{S}_g$ are concurrently input into the SCA module for context modeling in scene coupling. This process allows the SCA module to integrate and refine semantic information across different spatial contexts, addressing the intrinsic spatial correlation, thereby producing a semantically enhanced representation $\mathcal{R}_a$ as:
\begin{equation}
    \mathcal{R}_a = \mathrm{SCA}(\mathcal{R}_l, \mathcal{S}_l, \mathcal{S}_g).
\end{equation}
Finally, $\mathcal{R}_a$ is spatially restored and concatenated with $\mathcal{R}$ to produce the final output representation $\mathcal{R}_o$ which the discriminative properties of classes of pixel features are enhanced. Therefore, the pixel-level classification of various geospatial objects can be accomplished more accurately.

\subsection{Scene Coupling Attention Module}

This section introduces our proposed scene coupling which reconstructs the attention affinity operation to improve the attention modeling process. Specially, it embeds the scene information during attention computation, facilitating to leverage the inherent spatial correlations between geospatial objects in remote sensing images, thereby improving the model's segmentation performance. As shown in Figure \ref{attn}, this module decomposes the given scene information into two key elements: scene global representation $f$ and scene object distribution $r$. 

%This section will explain the computation principles of both elements.

\subsubsection{Scene Object Distribution}

\begin{figure*}[t]
    \centering \includegraphics[width=0.95\textwidth]{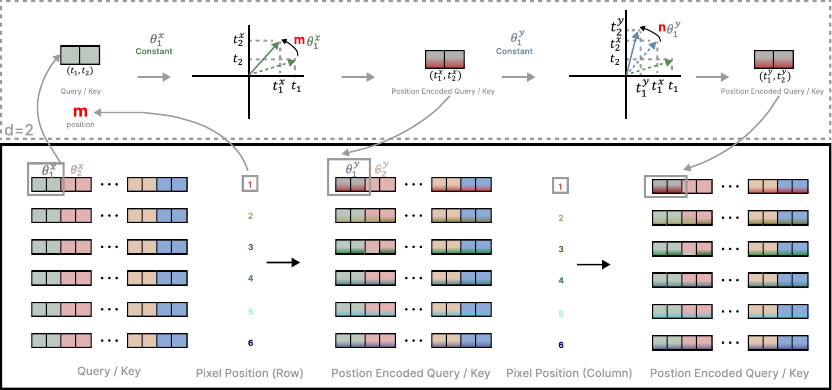}
    \caption{ROPE+ working analysis. We set the basic angles of rotation $\theta_i^x$ and $\theta_i^y$ in the width and height directions of different channels, respectively. Then, for a pixel feature with position $(m, n)$, we rotate it twice consecutively, each time with angles positively correlating with its position on the width and height, i.e., $m\theta_i^x$ and n$\theta_i^y$. Therefore, after two rotations, the transformed feature is able to encode the absolute positional information of the object, and indirectly encodes the relative distribution of the object in the subsequent process of attention affinity.}
    \label{scm}
\end{figure*}
%%%2024/07/06： 这个章节的公式很多，但是符号解释的不太清楚，一些计算过程不够清晰。

In remote sensing images, geospatial objects typically exhibit regular spatial distribution patterns. Particularly, neighboring objects usually appear in some form of combination or concurrency \cite{loveda}, and pixels near an object may show a high degree of correlation. For example, vehicles are typically found on roads rather than in fields, and buildings are usually located alongside roads but away from forests. Therefore, modeling the distribution of scene objects is crucial to encourage the model to learn spatial distribution patterns specific to certain scenes.

Inspired by ROPE's modeling of textual position information \cite{rope} by applying a rotation matrix representing the position, we propose a novel ROPE+ module, which indirectly models the relative distribution of targets within a scene through inner product of their absolute position information, and exhibits better generalization capabilities without the need for a fixed positional encoding length \cite{relative}. In this sense, we first define a 2D case of the original ROPE as:
\begin{equation}
\label{eq-2d}
\begin{aligned}
f_q(\boldsymbol{x}_m,m)& =(\boldsymbol{W}_q\boldsymbol{x}_m)e^{im\theta}, \\
f_k(\boldsymbol{x}_n,n)& =(\boldsymbol{W}_k\boldsymbol{x}_n)e^{in\theta}, \\
g(\boldsymbol{x}_m,\boldsymbol{x}_n,m-n)& =\mathrm{Re}[(\boldsymbol{W}_q\boldsymbol{x}_m)(\boldsymbol{W}_k\boldsymbol{x}_n)^*e^{i(m-n)\theta}] ,
\end{aligned}  
\end{equation}
where $\mathrm{Re}[\cdot]$ is the real part of a complex number and ${\boldsymbol{W}_k\boldsymbol{x}_n}^*$
represents the conjugate complex number of $\boldsymbol{W}_k\boldsymbol{x}_n$. $\theta \in \mathbb{R}$ is a pre-set non-zero constant, $\boldsymbol{x}_m$ and $\boldsymbol{x}_n$ denote the token at the m-th and n-th positions of the sequence $\boldsymbol{x}$, respectively. Here, the $f_{q,k}$ can be further written in a multiplication matrix as:
\begin{equation}
\begin{split}
    \left.f_{\{q,k\}}(\boldsymbol{x}_m,m)=\left(\begin{array}{cc}\cos m\theta&-\sin m\theta\\\sin m\theta&\cos m\theta\end{array}\right.\right) \otimes\\
    \left(\begin{array}{cc}W_{\{q,k\}}^{(11)}&W_{\{q,k\}}^{(12)}\\W_{\{q,k\}}^{(21)}&W_{\{q,k\}}^{(22)}\end{array}\right)\left(\begin{array}{c}x_m^{(1)}\\x_m^{(2)}\end{array}\right)     .
\end{split}
\end{equation}
Obviously, the original RoPE encodes absolute positions using a rotation matrix while incorporating explicit relative position dependence in the self-attention formulation. This property facilitates modeling complex object distributions in remote sensing scenes that do not need to be constrained by the input resolution of the image.

Since the original Rope only applicable for unidirectional sequences such as text, we extend the original RoPE to the image domain by simultaneously imposing a rotation matrix with reference to the width and height of the target pixel as:
\begin{equation}
\begin{aligned}
f_q(\boldsymbol{x}_{(m^x,m^y)},m^x, m^y)& =(\boldsymbol{W}_q\boldsymbol{x}_m)e^{im^x\theta^x}e^{im^y\theta^y} \\
&= (\boldsymbol{W}_q\boldsymbol{x}_m)e^{i(m^x\theta^x+m^y\theta^y)}, \\
f_k(\boldsymbol{x}_{(n^x,n^y)},n^x,n^y)& =(\boldsymbol{W}_q\boldsymbol{x}_n)e^{in^x\theta^x}e^{in^y\theta^y} \\
&=(\boldsymbol{W}_k\boldsymbol{x}_n)e^{i(n^x\theta^x+n^y\theta^y)}, \end{aligned}  
\end{equation}
where the inner product between the rotated query \\
$f_q(\boldsymbol{x}_{(m^x,m^y)},m^x, m^y)$ and key $f_k(\boldsymbol{x}_{(n^x,n^y)},n^x,n^y)$ is calculated as:
\begin{equation}
\begin{aligned}
g(\boldsymbol{x}_{(m^x,m^y)},\boldsymbol{x}_{(n^x,n^y)},m^x-n^x,m^y-n^y) = \\
\mathrm{Re}[(\boldsymbol{W}_q\boldsymbol{x}_m)(\boldsymbol{W}_k\boldsymbol{x}_n)^*e^{i((m^x-n^x)\theta^x+(m^y-n^y)\theta^y)}] .
\end{aligned}  
\end{equation}
Consequently, we extend the RoPE from 2D cases to higher dimensions to reconstruct Eq. \ref{eq-2d},
\begin{equation}
\begin{split}
    \breve{f}_{\{q,k\}}(\boldsymbol{x}_{(m^x,m^y)},m^x, m^y)=(R_{(\Theta^x,m^x)}^d&R_{(\Theta^y,m^y)}^d)\otimes \\
    &W_{\{q,k\}}\boldsymbol{x}_{(m^x,m^y)} .
\end{split}
\end{equation}

Then, the rotation matrix $\boldsymbol{R}_{\Theta,m}^d$ of the width direction and the height direction are both computed as,
\begin{equation}
\scriptsize
\boldsymbol{R}_{\Theta,m}^d=\begin{pmatrix}\cos m\theta_1&-\sin m\theta_1&0&\cdots&0&0\\\sin m\theta_1&\cos m\theta_1&0&\cdots&0&0\\0&0&\cos m\theta_2&\cdots&0&0\\0&0&\sin m\theta_2&\cdots&0&0\\\vdots&\vdots&\vdots&\vdots&\ddots&\vdots\\0&0&0&\cdots&\cos m\theta_{d/2}&-\sin m\theta_{d/2}\\0&0&0&\cdots&\sin m\theta_{d/2}&\cos m\theta_{d/2}\end{pmatrix}.
\end{equation}
To ensure that the model has different sensitivities in the width and height directions thus enhance the awareness of the location of geospatial objects, we define the corresponding basic rotation angles $\Theta^x$ (width direction) and $\Theta^y$ (height direction) as,
\begin{equation}
    \Theta^x = \{\theta_i^x=10000^{-2(i-1)/d},i \in [1,2,\dots, d/2]\},
\end{equation}
\begin{equation}
    \Theta^y = \{\theta_j^y=10000^{-2(i-1)+1/d},j \in [1,2,\dots, d/2]\}.
\end{equation}
Since processing of features in the complex domain is introduced, every two adjacent channels are considered as a pair of channels in the complex domain, representing the real and imaginary parts of the feature values, respectively. Therefore, it can be understood that $\Theta^x_i$ denotes the rotational base angle of the channel in the $i_\text{th}$ pair of complex domains in the horizontal direction, and $\Theta^y_j$ denotes the rotational base angle of the channel in the $j_\text{th}$ pair of complex domains in the vertical direction. Thus, the inner product of the two in the higher dimensional state is computed as:
\begin{equation}
\begin{split}
\boldsymbol{q}_m^\intercal\boldsymbol{k}_n=&(R_{(\Theta^x,m^x)}^dR_{(\Theta^y,m^y)}^d\boldsymbol{W}_q\boldsymbol{x}_{(m^x,m^y)})^\intercal\otimes \\
&(R_{(\Theta^x,n^x)}^dR_{(\Theta^y,n^y)}^d\boldsymbol{W}_k\boldsymbol{x}_{(n^x,n^y)}) \\
=&\boldsymbol{x}_{(m^x,m^y)}^\intercal\boldsymbol{W}_qR_{\Theta^x,n^x-m^x}^dR_{\Theta^y,n^y-m^y}^d\boldsymbol{W}_k\boldsymbol{x}_{(n^x,n^y)}.
\end{split}
\end{equation}
Thus, the Equation \ref{eq-begin-3} can be reconstructed as:
\begin{equation}
\label{eq-final-1}
   t_{m,n}=\frac{{\boldsymbol{q}^\intercal_m} \boldsymbol{k}_n}{\sqrt{C}} = \frac{{\Breve{f}_q(\boldsymbol{x})}_m^\intercal{\Breve{f}_k(\boldsymbol{x})}_n}{\sqrt{C}}.
\end{equation}
This way, through attentional interactions in Eq. \ref{eq-final-1}, the distribution of objects within the scene (i.e., $n^x-m^x$ and $n^y-m^y$) can be modelled, thus improving the model's representation of the segmentation of the feature objects.

\subsubsection{Scene Global Representation}

\begin{figure}[t]
	\centering	\includegraphics[width=0.5\textwidth]{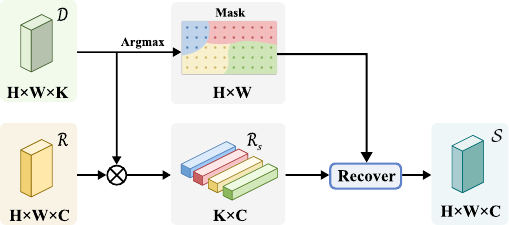}
	\caption{Semantic Mask Generation Module utilizes pre-classification masks for class-level contextual modeling of features, mitigating noise interference caused by a large number of background pixels.}
	\label{sm}
\end{figure}

Due to differences and complexity of remote sensing image scenarios, the same object may show varied relationships with objects adjacent to it under different scenes. For instance, rivers in urban areas are often found alongside roads, whereas in rural areas, rivers are surrounded by farmlands. This observation suggests that embedding a global representation (i.e., the context) of the scene may help models to learn more accurate relationships of geospatial objects. Moreover, attention mechanisms inherently model contextual relationships in the spatial domain, yet they lack exploration in the frequency domain. Inspired by \cite{fcanet,aff} that they fuse token information more efficiently based on the Fourier transform or discrete cosine transform, this paper propose to model the global representation of the scene from the frequency domain, aiming to capture valuable spectral information that is challenging to be observed in the spatial domain. 

\setlength{\tabcolsep}{10pt}
\begin{table*}[t]
	\begin{center}
	\caption{Comparison with state-of-the-art methods on the test set of the LoveDA dataset. Please note that the LoveDA dataset requires an online test to evaluate the model. Therefore results for the F1 and mAcc metrics are not available here. Per-class best performance is marked in bold, and the second largest value is underlined.}
	\label{table-loveda}
	\begin{tabular}{l||ccccccc||c}
	\Xhline{1.2pt}
            \rowcolor{mygray}
		     &\multicolumn{7}{c||}{IoU} &\\
            \rowcolor{mygray}
            \multicolumn{1}{c||}{\multirow{-2}{*}{Method}}&Background &Buildings &Roads &Water &Barren &Forest &Farmland &\multirow{-2}{*}{mIoU}\\
			
                \hline \hline
                PSPNet~\cite{pspnet}  & 44.4 & 52.1 &53.5 &76.5 &9.7 &44.1 &57.9 &48.3\\
			DeepLabv3+~\cite{deeplabv3+} & 43.0 & 50.9 &52.0 &74.4 &10.4 &44.2 &58.5 &47.6\\
			DANet~\cite{danet} &44.8 &55.5 &53.0 &75.5 &17.6 &45.1 &60.1 &50.2 \\
			Semantic FPN~\cite{fpn} &42.9 & 51.5 &53.4 &74.7 &11.2 &44.6 &58.7 &48.2\\
			FarSeg~\cite{farseg} &43.1 &51.5 &53.9 &76.6 &9.8 &43.3 &58.9 &48.2 \\
			OCRNet~\cite{ocrnet} &44.2 &55.1 &53.5 &74.3 &18.5 &43.0 &60.5 &49.9 \\
			LANet~\cite{lanet}  &40.0 &50.6 &51.1 &78.0 &13.0 &43.2 &56.9 &47.6 \\
			ISNet~\cite{isnet}&44.4 &57.4&58.0 &77.5 &\bf21.8 &43.9 &60.6 &51.9\\
			Segmenter~\cite{segmenter} &38.0 &50.7 &48.7 &77.4 &13.3 &43.5 &58.2 &47.1\\
			SwinUperNet~\cite{swintransformer} &43.3 &54.3 &54.3 &78.7 &14.9 &45.3 &59.6 &50.0\\
			MANet~\cite{manet} &38.7 &51.7 &42.6 &72.0 &15.3 &42.1 &57.7 &45.7 \\
                FLANet~\cite{flanet} & 44.6 & 51.8 &53.0 &74.1 &15.8 &45.8 &57.6 &49.0 \\
		      ConvNeXt~\cite{convnext}  & 46.9 & 53.5 &56.8 &76.1 &15.9 &47.5 &61.8 &51.2\\
			PoolFormer~\cite{poolformer} & 45.8 & 57.1 &53.3 &\bf80.7 &\underline{19.8} &45.6 &64.5 &52.4\\
                   BiFormer~\cite{biformer} &43.6 &55.3 &55.9 &79.5 &16.9 &45.4 &61.5 &51.2\\
      EfficientViT\cite{efficientvit}&42.9 &51.0 &52.8 &75.7 &4.3 &42.0 &61.2 &47.1\\
DDP \cite{ddp}&46.2 &57.2 &\underline{58.2} &\underline{80.3} &14.9 &46.5 &64.3 &52.5\\
LOGCAN++\cite{logcan++}&\underline{47.4} &\underline{58.4} &56.5 &80.1 &18.4&\textbf{47.9} &\underline{64.8} &\underline{53.4}\\
\hline
			SCSM &\bf48.3	&\bf60.4	&\bf58.4	&\bf80.7	&19.6	&\underline{47.6}	&\bf67.2 &\bf54.6\\
			\Xhline{1.1pt}
		\end{tabular}
	\end{center}
\end{table*}
\setlength{\tabcolsep}{2pt}

We first derive the global scene representation $\mathcal{G}$ using the Discrete Cosine Transform (DCT). For ease of subsequent analysis, this section first presents the expression for the two-dimensional DCT basis functions as:
\begin{equation}
    B_{h, w}^{x, y}=\alpha_{h}\alpha_{w}\cos\left(\frac{\pi(2x+1)h}{2T_H}\right)\cos\left(\frac{\pi(2y+1)w}{2T_W}\right),
\end{equation}
where $T_H$ and $T_W$ denote the height and width of the transformed region respectively, and
\begin{equation}
\alpha_{h}=\begin{cases}
    {1/\sqrt{T_H},} & {h=0}\\
    {\sqrt{2/T_H},} & {1 \leq h \leq T_H-1,}
\end{cases}
\end{equation}
\begin{equation}
    \alpha_{w}=\begin{cases}
    {1/\sqrt{T_W},} & {w=0}\\
    {\sqrt{2/T_W},} & {1 \leq w \leq T_W-1.}
\end{cases}
\end{equation}
The two-dimensional DCT is computed as follows:
\begin{equation}
\label{eq-dct}
    \mathcal{T}_{h, w}=\sum_{x=0}^{T_H-1} \sum_{y=0}^{T_W-1} A_{x, y} B_{h, w}^{x, y},
\end{equation}
where $A \in \mathbb{R}^{T_H \times T_W}$ represents the input image and $\mathcal{T} \in \mathbb{R}^{T_H \times T_W}$ denotes the spectral domain of the two-dimensional DCT, $T_H$ and $T_W$ being the height and width of $A$, respectively. The inverse process of the two-dimensional DCT can be represented as:
\begin{equation}
    A_{x, y}=\sum_{h=0}^{T_H-1} \sum_{w=0}^{T_W-1} \mathcal{T}_{h, w} B_{h, w}^{x, y}.
\end{equation}

In the practice of this paper, we obtain the scene global representation at a certain frequency based on the Eq. \ref{eq-dct}, it can also be expressed as,
\begin{equation}
\label{eq-dct}
    \mathcal{T}_{u_i, v_i}=\sum_{x=0}^{T_H-1} \sum_{y=0}^{T_w-1} A_{x, y} B_{u_i, v_i}^{x, y},
\end{equation}
where $B_{u_i, v_i}^{x, y}$ denotes the basis of the pre-trained frequency $\mathcal{T}_{u_i, v_i}$. 

After a 2D DCT transformation, the image is converted into a frequency spectrum, where each frequency component represents a specific pattern of the given scene. For instance, low frequency components represent structural information, while high frequency components convey spatial details. Thus, selecting the appropriate frequencies would help to get a higher quality scene global representation.
%This paper adopts a frequency prior method. 
Specifically, we follow previous work \cite{fcanet} to adopt a frequency prior method, which applies pre-training on ImageNet \cite{imagenet}, keeping only one frequency variable at a time to determine the importance of each frequency variable. More details can be found in \cite{fcanet}. As a result, the top $M$ most task-specific frequency components are selected and concatenated along the channel dimension to build the global scene representation:
\begin{equation}
\mathcal{G} = \mathcal{J}[\mathcal{T}_{u_1,v_1}, \mathcal{T}_{u_2,v_2}, \dots, \mathcal{T}_{u_M,v_M}],
\end{equation}
where \(\mathcal{J}\) represents the concatenation operation along the channel dimension.

Finally, based on the obtained global scene representation \(f\), the paper reconstructs the Eq. \ref{eq-final-1} as:
\begin{equation}
\label{eq-final-2}
t_{m,n}=\frac{(\mathcal{G}\boldsymbol{q}^\intercal_m){\boldsymbol{k}_n}}{\sqrt{C}}=
\frac{\mathcal{G}{\Breve{f}_q(\boldsymbol{x})}_m^\intercal{\Breve{f}_k(\boldsymbol{x})}_n}{\sqrt{C}}.
\end{equation}
The structural diagram of the Scene Coupling Attention Module is illustrated in Figure \ref{scm}. It can be observed that our approach reconstructs the standard spatial attention module based on the global scene representation \(f\) and the scene object distribution \(r\), which can effectively capture the inherent spatial correlation between features during attention computation. %Moreover, the proposed Scene Coupling Attention Module has a clear mathematical formulation.

\begin{figure*}[t]
	\centering \includegraphics[width=0.9\textwidth]
       {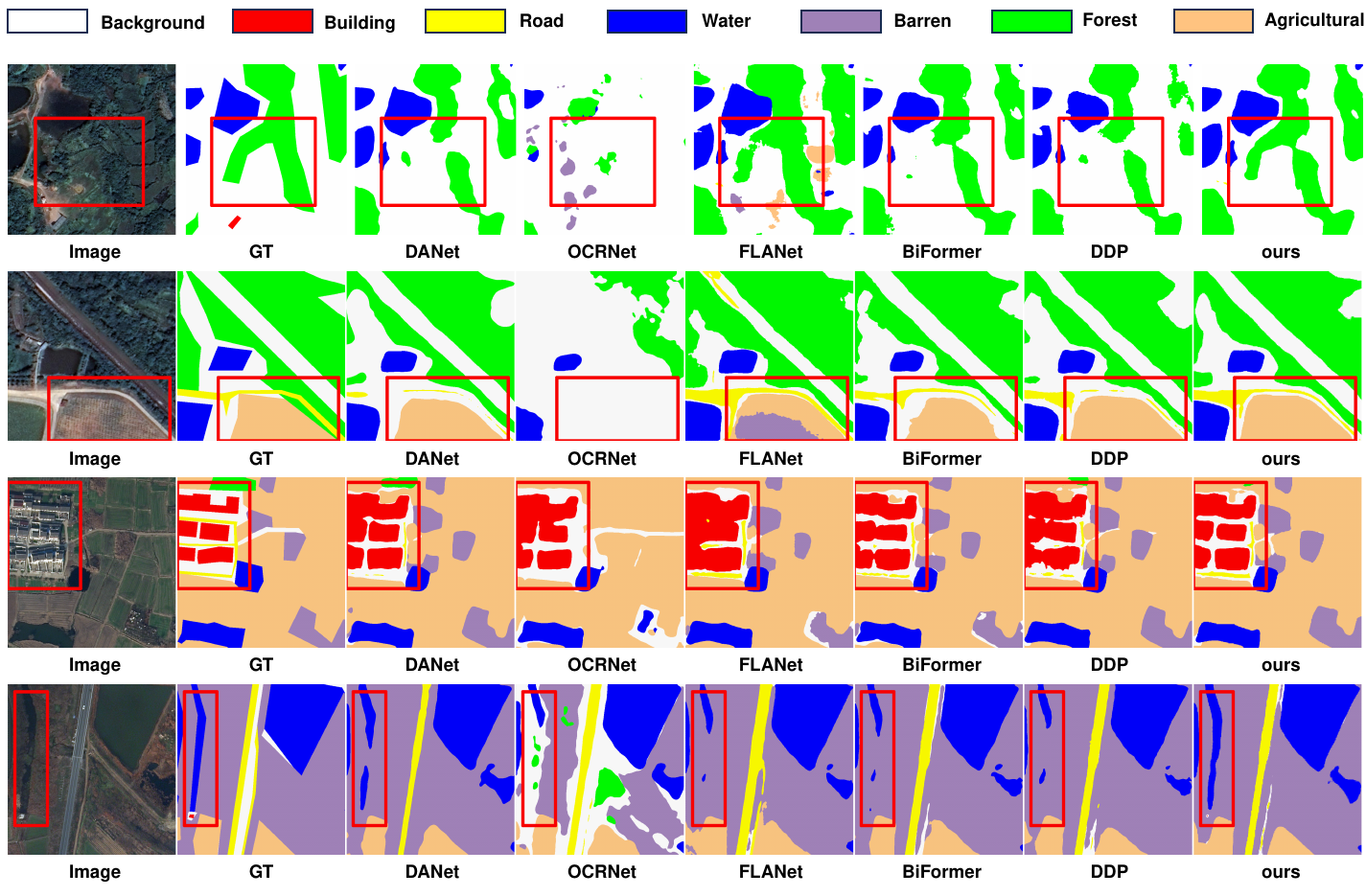}%
	\centering
	\caption{Qualitative comparison between SCSM and other state-of-the-art methods on the LoveDA test set. The red dashed box is the area of focus. Best viewed in color and zoom in.}
\label{fig:loveda}
\end{figure*}

\subsection{Semantic Mask Generation Module}

Remote sensing images are characterized by complex backgrounds and high intra-class variance, causing prior dense affinity-based spatial attention methods \cite{danet,ccnet,flanet} to introduce a large amount of background noises, thereby reducing algorithm performance. To avoid this issue, some class attention-based methods \cite{acfnet,ocrnet,ccanet} attempt to mitigate background interference by constructing representative class centers. However, they do not account for intra-class variance, where pixels in the feature space are far from the global class center, and thus would affect class-level context modeling. Moreover, these obtained class centers \cite{logcan++} lose spatial information and thus cannot effectively leverage the spatial correlation between targets in remote sensing image features. 

Consequently, this paper proposes a novel Local-Global Semantic Mask strategy. 
By using local semantic masking with spatial priors as intermediary perceptual elements to indirectly associate pixels with global semantic masks, it can accurate model class-level context, and combine scene coupling to produce performance gains between the two modules. As shown in Figure \ref{sm}, given a feature representation \(\mathcal{R}\in \mathbb{R} ^{\hat{C}\times H\times W}\) extracted by backbone, an initial classification operation (two consecutive \(1 \times 1\) convolutional layers) first generates the corresponding pre-classification representation \(\mathcal{D}\in \mathbb{R} ^{K\times H\times W}\), where \(K\) denotes the number of classes. The global class center \(\mathcal{S}\) is defined as:
\begin{equation}
\begin{aligned}
\mathcal{M}_D &= \mathrm{Argmax}_K(\mathcal{D}^{K\times H\times W}), \\
\mathcal{S} &= \psi(\mathcal{D}^{K\times (H\times W)}\otimes \mathcal{R}^{(H\times W)\times \hat{C}}, \mathcal{M}_D),
\end{aligned}
\end{equation}
where $\mathcal{M}_D^{H\times W}$ denotes the pre-classified mask, and \(\psi\) is a recover function that places the corresponding class centers on the position of original feature map (i.e., $\mathcal{R}^{H\times W\times \hat{C}}$) with the guidance of $\mathcal{M}_D$. Thus, we reassign spatial information (i.e., spatial prior) to the class centers to obtain the $\mathcal{S} ^ {H \times W \times \hat{C}}$, which is significantly different from previous work such as OCRNet \cite{ocrnet}, and LOGCAN++ \cite{logcan++}. These class centers with spatial prior can then be used to input scene-coupling attention modules.

Then, as shown in Fig. \ref{scsm}, \(\mathcal{R}\) and \(\mathcal{D}\) are segmented along the spatial dimension to obtain \(\mathcal{R}_l\) and \(\mathcal{D}_l\), respectively. Here, the local class representation \(\mathcal{S}_l\) is calculated as follows:
\begin{equation}
\begin{aligned}
\mathcal{M}_{D_l} &= \mathrm{Argmax}_K(\mathcal{D}_l^{(N_h \times N_w) \times K \times h \times w}), \\
\mathcal{S}_l &= \psi(\mathcal{D}_l^{(N_h \times N_w) \times K \times (h \times w)}\otimes \mathcal{R}_l^{(N_h \times N_w) \times (h \times w) \times \hat{C}}, \mathcal{M}_{D_l}),
\end{aligned}
\end{equation}
where \(h\) and \(w\) are the height and width of the selected local patch, and \(N_h = \frac{H}{h}\) and \(N_w = \frac{W}{w}\). Similarly, \(\mathcal{S}\) is segmented along the spatial dimension to obtain \(\mathcal{S}_g \in \mathbb{R}^{(N_h \times N_w) \times (h \times w) \times \hat{C}}\). As a result, we can reconstruct the Equation \ref{eq-begin-2} as:
\begin{equation}
\label{eq-final-3}
\boldsymbol{x^q} = \mathcal{R}_l,\quad \boldsymbol{x^k} = \mathcal{S}_l,\quad \boldsymbol{x^v} = \mathcal{S}_g.
\end{equation}

By modifying the vanilla attention operation, our approach ingeniously combines the local-global class attention designed with scene awareness and local-global class attention, which reduces the background noise interference, mitigates the damage of large intra-class variance for context modeling and effectively exploits the intrinsic spatial correlation of geospatial objects. Additionally, the proposed split operation can largely reduces the parameter and computation overhead, to keep the model lightweight.

\subsection{Loss Function}

This paper employs the standard cross-entropy loss to guide the training of SCSM module. Given a predicted mask \(\hat{y}\) (i.e., obtained after the softmax) and the corresponding ground truth mask \(y\) (i.e., obtained after the one-hot encoding), the cross-entropy loss is calculated as:
\begin{equation}
    \mathcal{L}_\text{ce} = -\frac{1}{HW}\sum_{i=0}^{HW-1}y_i \log(\hat{y}_i).
\end{equation}
The paper applies cross-entropy loss to supervise the training process of the SCSM. First, following previous work \cite{pspnet,deeplabv3+}, an FCN branch is extracted from the second-to-last residual block of the ResNet to compute auxiliary loss, which is named \(\mathcal{L}_\text{ce}^{a}\). This auxiliary loss helps constrain feature generation. Second, cross-entropy loss is added to the pre-classification representation \(\mathcal{D}\) and the final predicted mask, respectively named \(\mathcal{L}_\text{ce}^{d}\) and \(\mathcal{L}_\text{ce}^{o}\), to supervise mask generation. The final loss function for the model is:
\begin{equation}
    \mathcal{L} = \mathcal{L}_\text{ce}^{o} + 0.8\mathcal{L}_\text{ce}^{d} + 0.4\mathcal{L}_\text{ce}^{a}.
\end{equation}
It's important to note that, following previous work \cite{pspnet,deeplabv3+}, the coefficient for the auxiliary loss \(\mathcal{L}_\text{ce}^{a}\) is set to 0.4. For the pre-classification loss \(\mathcal{L}_\text{ce}^{d}\), an ablation study was conducted, and the best segmentation performance was achieved when the coefficient was set to 0.8.

\setlength{\tabcolsep}{18pt}
\begin{table*}[t]
    \begin{center}
        \caption{
        Comparative Results on the ISPRS Vaihingen and ISPRS Potsdam Datasets. Per-class best performance is marked in bold, and the second largest value is underlined.
        }
        \label{table-isprs}
        \begin{tabular}{l||ccc||ccc}
        \Xhline{1.2pt}
            \rowcolor{mygray}
             &\multicolumn{3}{c||}{ISPRS Vaihingen} &\multicolumn{3}{c}{ISPRS Potsdam}\\
            \rowcolor{mygray}
            \multicolumn{1}{c||}{\multirow{-2}{*}{Method}}&AF &mIoU &OA &AF &mIoU &OA\\
            
                \hline \hline
                PSPNet~\cite{pspnet}&86.47 &76.78 &89.36 & 89.98 &81.99 &90.14\\
            DeepLabv3+~\cite{deeplabv3+}&86.77 &77.13 &89.12 &90.86 &84.24 &89.18\\
            DANet~\cite{danet}&86.88 &77.32 &89.47 &89.60 &81.40 &89.73\\
            Semantic FPN~\cite{fpn}&87.58 &77.94 &89.86 & 91.53 &84.57 &90.16\\
            FarSeg~\cite{farseg}&87.88 &79.14 &89.57  &91.21 & 84.36 &89.87\\
            OCRNet~\cite{ocrnet} &89.22 &81.71 &90.47 &92.25 &86.14 &90.03\\
            LANet~\cite{lanet} &88.09 &79.28 &89.83 &91.95 &85.15 &90.84\\
            ISNet~\cite{isnet}&90.19 &82.36 &90.52 &92.67 &86.58 &91.27 \\
            Segmenter~\cite{segmenter}&88.23 &79.44 &89.93 &92.27 &86.48 &91.04\\
            SwinUperNet~\cite{swintransformer}&89.9 &81.8 &91.0 &92.24 &86.37 &90.98\\
            MANet~\cite{manet}&90.41 &82.71 &90.96 &92.90 &86.95 &91.32\\
                FLANet~\cite{flanet}&87.44 &78.08 &89.60 &\underline{93.12} &87.50 &\underline{91.87}\\
              ConvNeXt~\cite{convnext}&90.50 &82.87 &91.36 & 93.03 &87.17 &91.66\\
            PoolFormer~\cite{poolformer}&89.59 &81.35 &90.30 &92.62 &86.45 &91.12\\
           BiFormer~\cite{biformer}&89.65 &81.50 &90.63 &91.47 &84.51 &90.17\\
            EfficientViT~\cite{efficientvit}&87.56 &80.52 &89.41 &90.11 &84.23 &89.61\\
            DDP~\cite{ddp}&90.23 &82.56 &91.13 &93.05 &87.41 &91.76\\
            LOGCAN++ \cite{logcan++}&\underline{90.87} &\underline{83.89} &\underline{91.85} &93.11 &\underline{87.57} &91.48 \\
            \hline
            SCSM&\textbf{91.59} &\textbf{84.68} &\textbf{92.22} &\textbf{93.60} &\textbf{87.79} &\textbf{92.13}\\
            \Xhline{1.1pt}
        \end{tabular}
    \end{center}
\end{table*}
\setlength{\tabcolsep}{2pt}

\section{Experimental settings}
\subsection{Datasets}
To evaluate the segmentation performance of the model, experiments were conducted on four widely used remote sensing image datasets.

The ISPRS Vaihingen dataset \cite{vaihingen} is a commonly used dataset for remote sensing image segmentation. It contains aerial images of the German town of Vaihingen along with corresponding ground truth labels. The dataset is provided by DLR (German Aerospace Center) and is used for the ISPRS (International Society for Photogrammetry and Remote Sensing) competitions. The ISPRS Vaihingen dataset includes 33 images with a ground sampling distance (GSD) of 9 cm, collected from a small village featuring numerous individual buildings and small multi-story structures. Each orthoimage includes three multispectral bands (near-infrared, red, green) along with a digital surface model (DSM) and a normalized digital surface model (NDSM). The dataset comprises six categories, namely impervious surfaces, buildings, low vegetation, trees, cars, and clutter/background. Note that only the red, green, and blue channels are used for experiments in this section. Image sizes vary from 1996 × 1995 pixels to 3816 × 2550 pixels. In this paper, we utilize 16 images for training, namely: area\_1, area\_3, area\_5, area\_7, area\_11, area\_13, area\_15, area\_17, area\_21, area\_23, area\_26, area\_28, area\_30, area\_32, area\_34, and area\_37, with the remaining 17 images used for testing.

\begin{figure*}[t]
	\centering \includegraphics[width=0.9\textwidth]
       {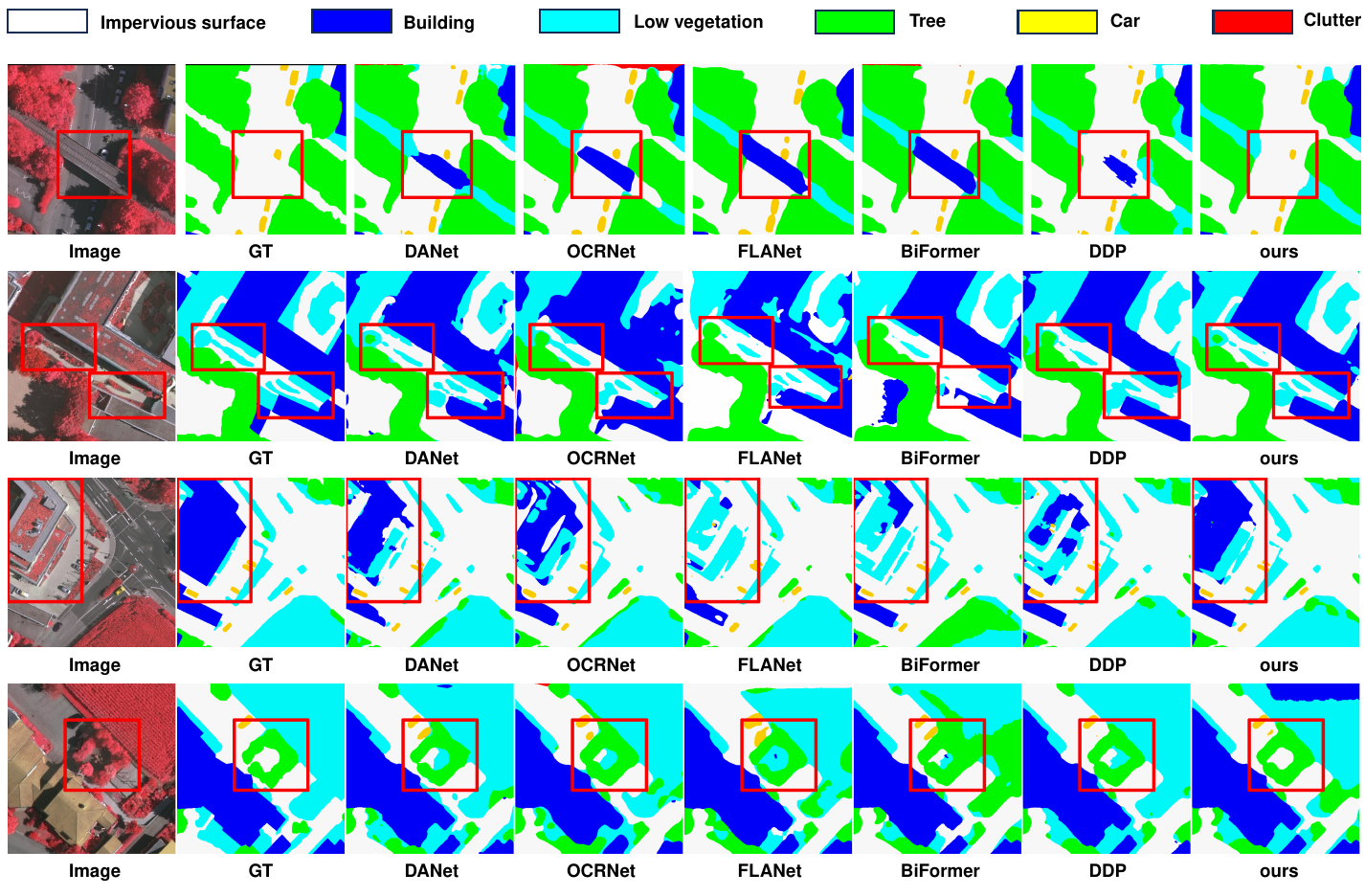}%
	\centering
	\caption{Qualitative comparison between SCSM and other state-of-the-art methods on the Vaihingen test set. The red dashed box is the area of focus. Best viewed in color and zoom in.}
\label{fig:vaihingen}
\end{figure*}

The ISPRS Potsdam dataset \cite{vaihingen} is another commonly utilized dataset for remote sensing image segmentation, also provided by DLR. It includes aerial images of the historic city of Potsdam, Germany, along with corresponding ground truth labels. Comprising 38 images with a ground sampling distance (GSD) of 5 cm, each image measures 6000 x 6000 pixels. The imagery captures a historic urban area characterized by large building blocks, narrow streets, and dense settlement structures. The dataset features four multispectral bands (i.e., near-infrared, red, green, blue) along with digital surface models (DSMs) and normalized digital surface models (NDSMs), including the same category labels as found in the ISPRS Vaihingen dataset. Similarly, only the red, green, and blue channels are utilized for experiments. In this paper, we employ 24 images for training, specifically: area\_2\_10, area\_2\_11, area\_2\_12, area\_3\_10, area\_3\_11, area\_3\_12, area\_4\_10, area\_4\_11, area\_4\_12, area\_5\_10, area\_5\_11, area\_5\_12, area\_6\_7, area\_6\_8, area\_6\_9, area\_6\_10, area\_6\_11, area\_6\_12, area\_7\_7, area\_7\_8, area\_7\_9, area\_7\_10, area\_7\_11, and area\_7\_12, with the remaining 14 images used for testing.

The LoveDA dataset \cite{loveda} comprises 5,987 high-resolution optical remote sensing images of 1024 × 1024 pixels (GSD 0.3 m) from three different cities, encompassing seven land cover categories: buildings, roads, water, barren land, forests, agriculture, and background. Additionally, the LoveDA dataset includes two domains (urban and rural), posing significant challenges such as multi-scale targets, complex background samples, and inconsistent sample distribution. The paper utilizes 2522 images for training, 1669 images for validation, and the remaining 1796 images for testing.

The iSAID dataset \cite{isaid} consists of 2,806 remote sensing images sourced from multiple satellites and sensors, with original image sizes ranging from 800 × 800 pixels to 4000 × 13,000 pixels. In addition, iSAID stands as one of the largest geospatial semantic segmentation datasets for remote sensing imagery, containing 655,451 densely annotated object instances across 15 categories within 2,806 high-resolution images. The dataset is divided into predefined training, validation, and test sets, with 1411, 458, and 937 images respectively.

\subsection{Evaluation Metrics}
We use three common metrics include F1 score, mean Intersection over Union (mIoU), and Overall Accuracy (OA) to evaluate the segmentation performance of SCSM. Following prior work \cite{farseg,manet,pointflow}, this paper selects the F1 score as the primary evaluation metric for the ISPRS Vaihingen and ISPRS Potsdam datasets. For the iSAID and LoveDA datasets, mean IoU (mIoU) is chosen as the main evaluation criterion.

We first give the formulation of mIoU, i.e., 
\begin{equation}
    \rm mIoU = \sum_{k=1}^{K} \text{IoU}_k,
\end{equation}
where k denotes the category index and K is the total number of categories. IoU is the ratio of the intersection to the union of the predicted and actual segmentations, calculated by:
\begin{equation}
    \rm{IoU} = \frac{TP}{TP + FP + FN},
\end{equation}
where TP (True Positive) refers to the correct predictions of the positive class, FN (False Negative) refers to the misclassification of positive samples as negative, FP (False Positive) refers to the misclassification of negative samples as positive, and TN (True Negative) refers to the correct predictions of the negative class.

\setlength{\tabcolsep}{4pt}
\begin{table*}[t]
	\begin{center}
		\caption{Comparative Results on the iSAID Datasets. Per-class best performance is marked in bold, and the second largest value is underlined.}
		\label{table-isaid}
		\begin{tabular}{c||ccccccccccccccc||c}
		\Xhline{1.2pt}
            \rowcolor{mygray}
	    &\multicolumn{15}{c||}{IoU} & \\
            \rowcolor{mygray}
			\multicolumn{1}{c||}{\multirow{-2}{*}{Method}}&Ship &ST &BD &TC &BC &GTF &Br. &LV &SV &HC &SP &RA &SBF &Pl. &Ha. &\multirow{-2}{*}{mIoU}\\
			
                \hline \hline
                 FCN-8s\cite{fcn}&51.7 & 22.9 & 26.4 &74.8 & 30.2 &27.8 &8.1 & 49.3 & 37.0 & 0 & 30.7 &51.9 &52.0 &62.9 &42.0 &41.6\\
                SPGNet\cite{spgnet} & 53.1 &43.3 &59.1 &74.7 &48.5 &43.7 &11.4 &52.8 &31.0 &4.4 &39.4 &33.7 &59.9 &45.3 &45.8 &46.5 \\
                DenseASPP\cite{denseaspp} &61.1 &50.0 &67.5 &86.0 &56.5 &52.2 &29.6 &57.1 &38.4 &0 &43.2 &64.8 &74.1 &78.1 &51.0 &56.8 \\
                NonLocal\cite{nonlocal} & 63.4 &48.0 &49.5 &86.4 &62.7 & 50.0 &35.0 &57.7 &43.4 &31.6 &44.9 &67.4 &71.0 &80.0&51.5 &58.8 \\
                DeepLab v3\cite{deeplabv3} & 59.7 &50.4&76.9&84.2&57.9&59.5&32.8&54.8&33.7&31.2&44.7&66.0&72.1&75.8&45.6&59.0 \\
                Semantic FPN\cite{fpn} &63.6 &59.4&71.7 &86.6 &57.7 &51.6 &33.9&59.1&45.1 &0&46.4&68.7 &73.5&80.8&51.2&59.3\\
                DANet\cite{danet} & 63.9 &46.2&73.7 &85.7 &57.9 &48.2&33.5 &57.9&43.2&36.1&45.7 &67.2&69.2 &80.4&52.3&60.0 \\
                RefineNet &63.8 &58.5 &72.3 &85.2 &61.0 &52.7 &32.6 &58.2 &42.3 &22.9 &43.4 &65.6 &74.4 &79.8 &51.1 &60.2 \\
                PSPNet\cite{pspnet} &65.2 &52.1 &75.7 &85.5 &61.1 &60.1 &32.4 &58.0 &42.9 &10.8 &46.7 &68.6 &71.9&79.5 &54.2 &60.2\\
                UNet\cite{unet} &63.7 &52.5 &67.1 &87.1 &57.6 &49.5 &33.9 &59.2 &47.8 &29.9 &42.2 &70.2 &69.5 &82.0 &54.6 &60.4 \\
                CCNet\cite{ccnet} & 64.7 &52.8 &65.0 &86.6 &61.4 &49.8 &34.6 &57.8 &43.3 &35.7 &44.6 &67.7 &70.0 &80.6 &53.0 &60.4\\
                DNLNet\cite{dnlnet} & 63.7 &52.2 &72.6 &86.6 &61.7 &54.1 &34.2 &56.8 &42.7 &36.8 &43.4 &68.2 &71.3 &79.9 &50.7&60.8 \\
                GCNet\cite{gcnet} & 64.9 &49.8 &72.4 &85.8 &59.3 &51.1 & 34.1 &58.3 &43.5 &34.9 &46.7 &68.8 &72.6 &80.8 &53.2&60.9\\
                OCNet\cite{ocnet} & 65.2 &48.3 &71.8 &87.0 &57.2 &55.9 &31.2 &59.5 &43.5 &34.9 &47.9 &70.2 &72.8 &80.9 &50.6 &61.0\\
                EMANet\cite{emanet} & 65.3 &52.8 &72.2 &86.0 &62.8 &49.0 &34.9 &57.6 &43.1 &38.6 &46.0 &69.2 &69.3 &80.7 &52.8 &61.2 \\
                DeepLab v3+\cite{deeplabv3+} & 63.7 &58.4 &75.6 &86.5 &59.9 &58.6 &34.9 &59.1 &43.9 &27.9 &48.2 &68.7 &74.5 &80.3 &51.9 &61.9\\
                HRNet\cite{hrnet} & 67.2 &64.4 &78.2 &87.6 &60.9 &57.5 &34.8 &59.9 &47.7 &15.9 &48.9 &68.2 &74.5 &82.3 &57.0 &62.7 \\
                UperNet\cite{upernet} & 65.9 &59.0 &75.7&87.1&61.6&58.5&36.1&60.0&45.7&33.6&49.5&70.6&73.5&81.7&54.4&63.2\\
                SFNet\cite{sfnet} & 69.2 &68.3& 77.5 &87.5 &59.4 &55.1 &29.7 &60.3 &46.8 &29.3 &50.8 &71.0 &72.7 &82.9 &53.4 &63.3\\
                FarSeg\cite{farseg} &65.3 &61.8 &77.7 &86.3 &62.0 &56.7 &36.7 &60.5 &46.3 &35.8 &51.2 &71.3 &72.5 &82.0 &53.9&63.7 \\
                PFNet\cite{pointflow} & \underline{70.3} &\bf74.7 &77.8 &87.7 &62.2 &59.5 &\underline{45.2} &\underline{64.6} &50.2 &37.9 &50.1 &71.7 &\underline{75.4} &\underline{85.0} &\underline{59.3} &64.8 \\
                FarSeg++\cite{farseg++} & 67.6 &59.8 &75.0 &\underline{88.9}&\underline{66.7}&57.8&40.1&64.4&\underline{51.6}&38.5&52.0&73.2&74.3&84.9&57.0&\underline{65.7}\\
                LOGCAN++ \cite{logcan++} &68.4 &70.2 &\underline{78.8} &87.6 &66.3 &\underline{61.1} &44.3 &63.2 &50.8 &\underline{38.9} &\underline{52.9} &\underline{74.3} &74.7 &84.3 & 59.1& 65.0\\
                \hline
                SCSM &\bf71.4 &\underline{72.3} &\bf80.3 &\bf89.2 &\bf68.8 &\bf62.6 &\bf45.3 &\bf66.7 &\bf52.4 &\bf40.8 &\bf53.9  &\bf73.9&\bf78.7 &\bf86.6 &\bf60.6 &\bf66.9\\
			\Xhline{1.1pt}
		\end{tabular}
	\end{center}
\end{table*}

F1-score (F1) is a metric that considers both the precision and recall of the model, which reflects both the accuracy and completeness of the model,
\begin{equation}
    \rm{F1} = \frac{2 \times Precision \times Recall}{Precision + Recall},
\end{equation}
where recision measures the proportion of true positive samples among all samples predicted as positive, calculated by:
\begin{equation}
    \rm{Precision} = \frac{TP}{TP + FP}.
\end{equation}
Recall measures the proportion of true positive samples correctly predicted as positive, given by:
\begin{equation}
    \rm{Recall} = \frac{TP}{TP + FN}.
\end{equation}
Similar to mIoU, the average F1-score (AF) across all categories is calculated by:
\begin{equation}
    \rm AF = \sum_{k=1}^{K} \text{F1}_k.
\end{equation}
Overall Accuracy (OA) measures the proportion of samples that are correctly predicted across all categories over the total number of samples, providing a straightforward metric of the model's overall classification accuracy:
\begin{equation}
    \rm{OA} = \frac{TP + TN}{TP + FN + FP + TN}.
\end{equation}

\begin{figure*}[t]
	\centering \includegraphics[width=1.0\textwidth]
       {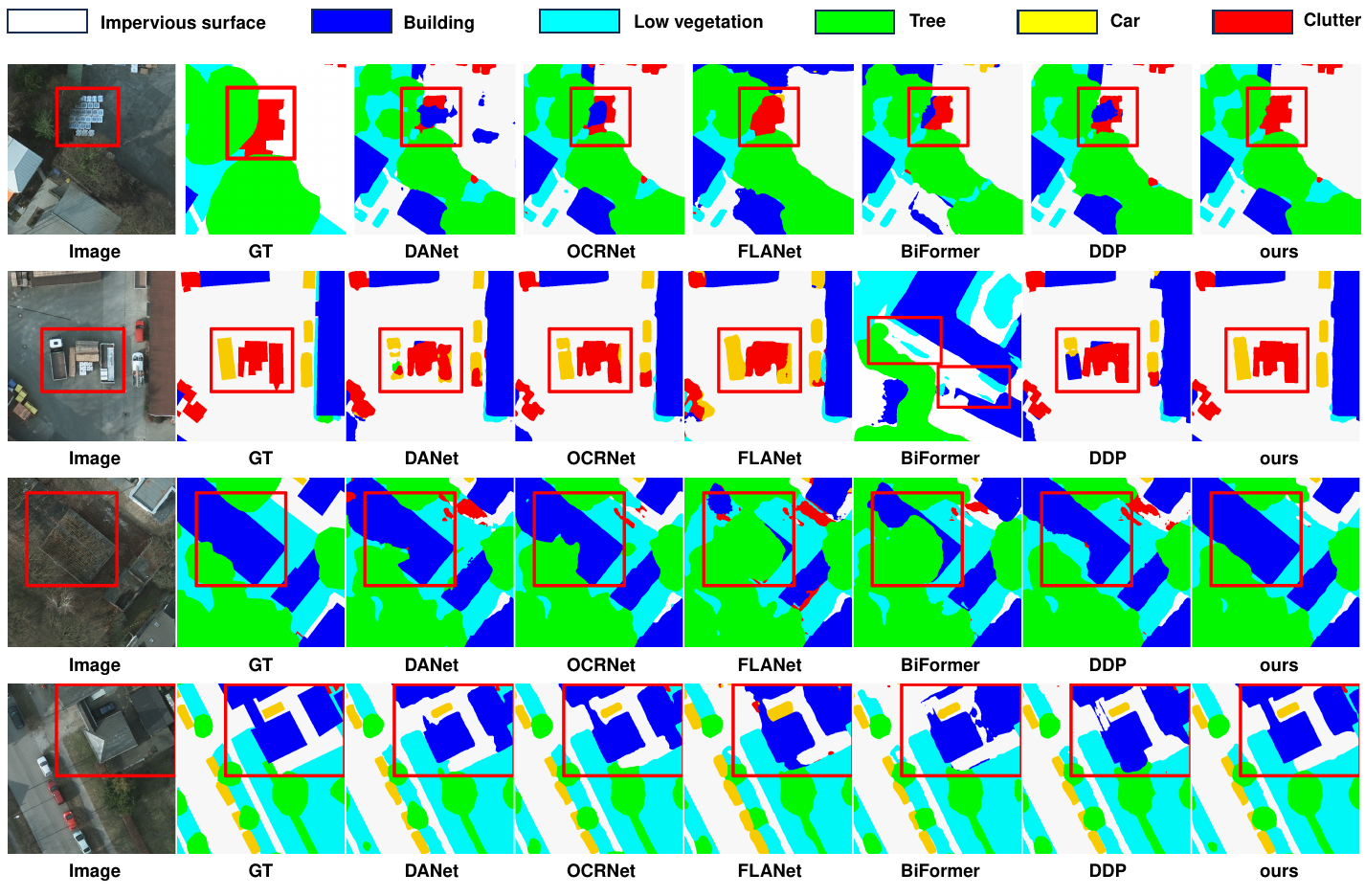}%
	\centering
	\caption{Qualitative comparison between SCSM and other state-of-the-art methods on the Potsdam test set. The red dashed box is the area of focus. Best viewed in color and zoom in.}
\label{fig:potsdam}
\end{figure*}

\subsection{Implementation Details}
The experiments in this section were conducted on a workstation equipped with eight NVIDIA Tesla V100 GPUs, each with 32 GB of VRAM, and were implemented using PyTorch. For the SCSM model, the SGD optimizer was utilized with an initial learning rate of 0.01 and a poly decay strategy for weight decay set at 0.0001. For the related comparison models, hyperparameters from their original publications were adhered to. Following previous work, random cropping was performed on the ISPRS Vaihingen, ISPRS Potsdam, LoveDA, and iSAID datasets, with cropping sizes of $512 \times 512$ for the first three datasets and $896 \times 896$ for the iSAID dataset. Additionally, data augmentation techniques were employed during training, including random scaling (scale factors of [0.5, 0.75, 1.0, 1.25, 1.5]), random vertical flipping, random horizontal flipping, and random rotation. For all four datasets, the total batch size was set at 16 and the total number of iterations at 80,000. A consistent experimental setup was used across all comparison methods to ensure fair comparisons.

\section{EXPERIMENTAL RESULTS AND ANALYSIS}
To validate the performance of the model, this section presents extensive comparative experiments. The comparison methods include approaches based on spatial context modeling such as PSPNet \cite{pspnet} and DeepLabv3+ \cite{deeplabv3+}; attention-based methods like DANet \cite{danet}, OCRNet \cite{ocrnet}, and ISNet \cite{isnet}; transformer-based approaches such as Segmenter \cite{segmenter}, PoolFormer \cite{poolformer}, BiFormer \cite{biformer} and EfficientViT \cite{efficientvit}; diffusion model-based models like DDP \cite{ddp}; along with classical remote sensing image segmentation methods like LANet \cite{lanet}, FarSeg \cite{farseg}, and PFNet \cite{pointflow}.
\subsection{Results on the ISPRS Loveda dataset}
\subsubsection{Qualitative analysis}
To validate the effectiveness of our SCSM, we first initially conducted experiments on the LoveDA dataset to evaluate the performance of the SCSM model. Thanks to the design of scene coupling and the local-global semantic mask, SCSM effectively handles both urban and rural scenes in the LoveDA dataset. Table~\ref{table-loveda} lists the specific comparison results. The best metric values are bolded, and the second-best values are underlined. Notably, SCSM achieved the highest mIoU at 54.6\%. Specifically, compared to the previously state-of-the-art method, PoolFormer \cite{poolformer}, SCSM shows an improvement of 2.2\% on the LoveDA dataset. Moreover, the performance improvements are particularly significant for common land cover targets such as buildings and farmlands. For instance, for buildings, SCSM shows a 3\% improvement over ISNet \cite{isnet}, and for farmlands, there is a 2.7\% improvement over PoolFormer \cite{poolformer}. These experimental results validate the effectiveness of the model.

\subsubsection{Qualitative analysis}
Fig. \ref{fig:loveda} visualizes the segmentation masks output by different models to qualitatively compare the segmentation performance of our SCSM with these competitors, where all input images come from the LoveDA test set. In the first three images, the output masks from our SCSM exhibit more complete shapes and clearer boundaries for forest, agriculture, and building. In the fourth image, our SCSM accurately identifies the water, whereas other methods such as OCRNet and FlANet do not perform as well. These results indicate that our SCSM demonstrates superior semantic recognition capability and visualization performance.

\setlength{\tabcolsep}{9pt}
\begin{table*}[t]
	\begin{center}
		\caption{
		Ablation Study on the Frequency Count on the Loveda (left) and ISPRS Vaihingen (Right) Dataset. The best value in each column is bolded.
		}
		\label{table-frequency}
		\begin{tabular}{c||cccccccc||c}
		\Xhline{1.2pt}
            \rowcolor{mygray}
			Frequency Count&Background &Buildings &Roads &Water &Barren &Forest &Farmland &mIoU &mIoU\\
			
                \hline \hline
                1  & {47.3}    & 58.1          & 57.0          & {81.3}    & 14.8          & 46.3          & 65.3          & 52.9    &82.88      \\
4  & 47.2          & 58.6          & {58.6}    & 80.6          & 19.2          & 46.5          & 64.5          & 53.6       &83.36   \\
8  & 47.1          & {59.8}    & 56.5          & \textbf{81.4} & {19.3}    & 46.5          & {65.5}    & {53.7}  &84.27  \\
16 & \textbf{48.3} & \textbf{60.4} & 58.4          & 80.7          & \textbf{19.6} & \textbf{47.6} & \textbf{67.2} & \textbf{54.6} &\bf 84.68\\
32 & {47.3}    & 59.6          & \textbf{59.3} & 80.9          & 17.3          & {47.1}    & 63.9          & 53.6        &84.16 \\
			\Xhline{1.1pt}
		\end{tabular}
	\end{center}
\end{table*}

\setlength{\tabcolsep}{8pt}
\begin{table*}
\begin{minipage}{\linewidth}
    	\centering
		\caption{
  Ablation Study on the impact of block size on the Loveda (left) and ISPRS Vaihingen (right) Dataset. The best value in each column is bolded.
		}
		\label{table-size}
		\begin{tabular}{c||cccccccc||c}
		\Xhline{1.2pt}
            \rowcolor{mygray}
			Block size&Background &Buildings &Roads &Water &Barren &Forest &Farmland&mIoU &mIoU\\
			
                \hline \hline
                7 &45.7	&57.4	&58.1	&79.7	&16.7	&45.2	&62.1	&52.1 &83.21\\
		      14  &46.2	&59.5	&56.3	&\bf81.1	&16.1	&46.8	&64.5	&52.9& 84.37\\
			21 &\bf48.3	&\bf60.4	&\bf58.4	&80.7	&\bf19.6	&\bf47.6	&\bf67.2	&\bf54.6 &\bf84.68\\
			28 &47.1	&59.7	&58.1	&80.9	&17.3	&\bf47.6	&64.6	&53.6 &83.57\\
			\Xhline{1.1pt}
		\end{tabular}
  \vspace{8mm}
\end{minipage}
\begin{minipage}{\linewidth}
    \centering
		\caption{
  Ablation Study of the basic rotation angles in the horizontal and vertical directions on the LoveDA (left) 
 and ISPRS Vaihingen (right) Dataset. Identical and different means that we set the identical and different basic rotation angles horizontally and vertically, respectively. The best value in each column is bolded.
		}
		\label{angle}
		\begin{tabular}{c||cccccccc||c}
		\Xhline{1.2pt}
            \rowcolor{mygray}
			Rotation angle&Background &Buildings &Roads &Water &Barren &Forest &Farmland&mIoU &mIoU\\
			
                \hline \hline
                Identical &47.3	&58.6	&\bf58.7	&\bf81.4	&18.8	&46.3	&62.8	&53.4 &83.27\\
			Different &\bf48.3	&\bf60.4	&58.4	&80.7	&\bf19.6	&\bf47.6	&\bf67.2	&\bf54.6 &\bf 84.68\\
			\Xhline{1.1pt}
		\end{tabular}
  \vspace{3mm}
\end{minipage}
\end{table*}

\setlength{\tabcolsep}{1pt}
\begin{table}[t]
	\begin{center}
		\caption{Comparison of Efficiency Metrics for the SCSM Model
		}
		\label{table-flop}
		\begin{tabular}{c||ccc}
		\Xhline{1.2pt}
            \rowcolor{mygray}
			Module &Params (M) &FLOPs (G) &Memory (MB)\\
			\hline \hline
			PPM~\cite{pspnet} & 23.1 & 309.5 & 257 \\
			ASPP~\cite{deeplabv3+} & 15.1 & 503.0 & 284 \\
			DAB~\cite{danet} &23.9 & 392.2 & 1546\\
			OCR~\cite{ocrnet} &10.5 & 354.0 &\underline{202} \\
			PAM+AEM~\cite{lanet} &\underline{10.4} & 157.6 &489\\
			ILCM+SLCM~\cite{isnet} &11.0 &180.6 &638\\
   FLA~\cite{flanet} &11.5 &\underline{154.9} &645\\	
			\hline
			SMG+CCA (Ours) &\bf2.4 &\bf40.5 &\bf135\\
   \Xhline{1.1pt}
		\end{tabular}
	\end{center}
\end{table}

\subsection{Results on the ISPRS Vaihingen and Potsdam dataset}
\subsubsection{Qualitative analysis}
We conduct a comparative analysis with the current state-of-the-art methods on the ISPRS Vaihingen and ISPRS Potsdam datasets. As depicted in Table~\ref{table-isprs}, SCSM achieves the best results in AF, mIoU, and OA metrics on these datasets, significantly outperforming other CNN-based and transformer-based models. Specifically, on the ISPRS Vaihingen dataset, SCSM exhibits improvements of 1.09\% in AF and 1.81\% in mIoU compared to the currently leading model, ConvNext\cite{convnext}. On the ISPRS Potsdam dataset, in comparison to the state-of-the-art model FLANet, SCSM demonstrates increases of 0.48\% in AF and 0.29\% in mIoU.

\subsubsection{Qualitative analysis}
Fig. \ref{fig:vaihingen} compares the visual results of our method with other models on the Vaihingen test set. In the first image, our method accurately identifies small objects such as cars, whereas other methods mistakenly classify them as houses. Additionally, in the other images, SCSM demonstrates clearer object boundary segmentation compared to other methods. Fig. \ref{fig:potsdam} compares the visual results of our method with other models on the Potsdam test set. In the first image, despite significant tree occlusion, our model accurately identifies clutter, while most other models are misled by the interference. In the second image, even in the presence of various similar interferences, our model can accurately distinguish between clutter and cars, indicating its high inter-class differentiation capability. Finally, in other images, our model shows more effective segmentation of object boundaries when dealing with large objects.

\subsection{Results on the iSAID dataset}
We conduct extensive comparative experiments on the iSAID dataset, as depicted in Table~\ref{table-isaid}. Due to space constraints, the table lists abbreviations for each category alongside their corresponding IoU metrics. The categories, from left to right, are as follows: ships, storage tanks, baseball diamonds, tennis courts, basketball courts, athletic tracks, bridges, large vehicles, small vehicles, helicopters, swimming pools, roundabouts, soccer fields, airplanes, and harbors. It is observed that SCSM outperforms the state-of-the-art methods PFNet and FarSeg++ in the mIoU metric by 2.1\% and 1.2\%, respectively. Notably, significant improvements are seen in challenging categories such as large vehicles, small vehicles, and helicopters, further substantiating the effectiveness of the model design.

\subsection{Complexity Comparison}
To evaluate the efficiency of the model, the paper compares it with mainstream contextual aggregation modules, as shown in Table \ref{table-flop}. Evaluation metrics include the number of parameters (Params), floating-point operations per second (FLOPs), and memory consumption. The experimental results indicate that SCSM significantly reduces the number of parameters, computational complexity, and memory consumption compared to mainstream contextual aggregation modules. In particular, compared to the methods ISNet (with context aggregation modules ILCM and SLCM) and FLANet based on global attention, SCSM reduces parameter count by around 80\%, computational complexity by around 75\%, and memory consumption by around 78\%. This further validates the effectiveness of the SMG and CCA module design.

\subsection{Ablation Study}
A series of ablation experiments were conducted on the LoveDA dataset to achieve the optimal model structure design, as follows.

\subsubsection{Ablation Study on the frequency Counts} The paper selects a certain number of frequencies as global scene representations to guide the modeling of spatial attention mechanisms. As shown in Table~\ref{table-frequency}, the chosen number of frequencies (M) are 1, 4, 8, 16, and 32. The experimental results indicate that the model achieves the best segmentation performance when the frequency count is 16. Therefore, in the final model structure, 16 frequency values are chosen and concatenated along the channel dimension to obtain the global scene representation $\mathcal{G}$.

\setlength{\tabcolsep}{6pt}
\begin{table*}[t]
	\begin{center}
		\caption{
		Ablation Study on the Impact of Model Structure Variations on the Loveda Dataset. The first line is base. The best value in each column is bolded.
		}
		\label{table-all}
		\begin{tabular}{c||cccccccc}
		\Xhline{1.2pt}
            \rowcolor{mygray}
			Method&Background &Buildings &Roads &Water &Barren &Forest &Farmland &mIoU \\
			
                \hline \hline
                Base                & 43.8  & 57.8  & 52.5  & 77.5      & 21.0 & 46.8   & 56.7  & 50.8 \\
Base+GA             & 46.0  & 55.5  & 59.5  & 79.6      & 16.8 & 46.8   & 61.6  & 52.2\\
Base+SCA          & {47.7}    & 59.7          & {56.7}    & 80.9          & 18.3          & 47.4          & {65.9}    & 53.8          \\
Base+SMG(G+G)+GA  & 47.1          & 59.8          & 56.5          & \textbf{81.1} & 19.5          & \textbf{48.2} & 64.5          & 53.8          \\
Base+SMG(G+L)+GA  & 47.1          & 60.3          & 55.9          & \textbf{81.1} & \textbf{21.7} & 47.2          & {65.9}    & 54.2          \\
Base+SMG(G+G)+SCA & 47.2          & \textbf{60.5} & {56.7}    & {81.0}    & {20.8}    & {48.0}    & 65.7          & {54.3}  \\
SCSM             & \textbf{48.3} & {60.4}    & \textbf{58.4} & 80.7          & 19.6          & 47.6          & \textbf{67.2} & \textbf{54.6}\\
			\Xhline{1.1pt}
		\end{tabular}
	\end{center}
\end{table*}

\begin{figure*}[t]
	\centering \includegraphics[width=1.0\textwidth]
       {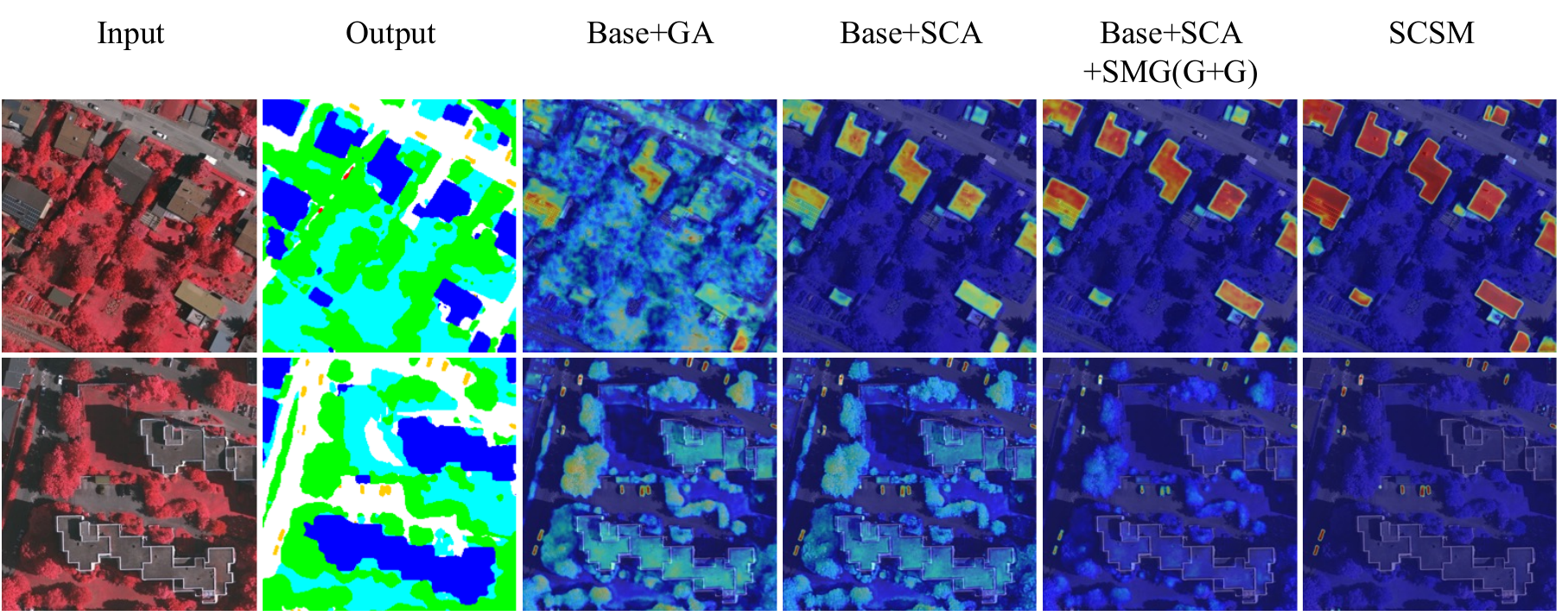}
	\centering
	\caption{Ablation Study on the Impact of Model Structure Variations with Class Activation Maps. The target activation classes are building (first line) and car (second line), respectively.}
\label{fig:ablation_cam}
\end{figure*}

\setlength{\tabcolsep}{6pt}
\begin{table*}[t]
	\begin{center}
		\caption{
		Ablation Study on the Impact of ROPE+ on the Loveda Dataset. The first line is base. The best value in each column is bolded.
		}
		\label{table-rope}
		\begin{tabular}{c||cccccccc}
		\Xhline{1.2pt}
            \rowcolor{mygray}
			Method&Background &Buildings &Roads &Water &Barren &Forest &Farmland &mIoU \\
SCSM & \textbf{48.3} & \bf 60.4 & \textbf{58.4} & \bf80.7  & \bf19.6          & \bf47.6   & \textbf{67.2} & \textbf{54.6}\\
No ROPE+ &47.3 &59.3 &57.4 &80.3 &19.1 &46.9 &66.4 &53.8\\
Sinusoidal \cite{detr} &47.0 &58.8 &57.1 &80.4 &17.2 &47.2 &66.4 &53.4\\
CPVT \cite{cpvt} &47.9 &60.2 &58.1 &80.3 &19.5 &46.9 &67.0&54.3\\
			\Xhline{1.1pt}
		\end{tabular}
	\end{center}
    \vspace{4mm}
\end{table*}

\subsubsection{Ablation Study on the Block Size}

Due to the introduction of spatial dimension block cutting operations, the size of the blocks has a significant impact on model performance, and this section explores this aspect. Specifically, the paper sets the block size \(P\) as 7, 14, 21, and 28. It is important to note that since the paper selects a frequency prior, namely, the \(M\) frequency values with strong responses obtained from a model pre-trained on ImageNet (where the final output size of the model is \(7 \times 7\)), the block size is set as a multiple of 7 to avoid interpolation operations and preserve the effectiveness of the pre-training. Furthermore, the paper employs overlapping block operations to ensure that the model can segment into an integer number of local blocks. As shown in Table~\ref{table-size}, the model achieves the best segmentation performance when the block size is set to 21.

\subsubsection{Ablation Study on the Rotation Angles}
We conduct experiments on the LoveDA dataset to explore the effectiveness of setting different basic rotation angles (i.e., $\Theta^x$ and $\Theta^y$) in the horizontal and vertical directions. The experimental results are shown in Table \ref{angle}, which validate our hypothesis. It can be understood that different basic rotation angles facilitate the model to possess different sensitivities to the distribution of geospatial objects in horizontal and vertical angles during the affine process. As a result, the model is able to mine more distribution pattern information from the images. In other words, the model has more information redundancy, and this redundancy facilitates the enhancement of the model's recognition capability \cite{ghostnet,ghostnetv2}.

\subsubsection{Ablation Study on the Overall Structure}
The paper conducts ablation experiments to validate the effectiveness of the SMG (Semantic Mask Generation) module and the SCA (Scene Context Attention) module, as shown in Table~\ref{table-all}. The SMG and SCA modules were removed, leaving only the backbone network and the FCN (Fully Convolutional Network) decoding head as the model's baseline (Base). Additionally, in the table, 'L' denotes Local semantic masks, and 'G' denotes Global semantic masks. Comparing the second and third rows of the table reveals that the SCA module contributes more significantly to remote sensing image segmentation than a general spatial attention (GA) module. Furthermore, comparing the fourth and fifth rows demonstrates the effectiveness of introducing local semantic masks as intermediary perceptual elements to indirectly link pixels with global semantic masks, enhancing segmentation performance. Therefore, the paper selects Base+SMG(G+L)+SCA as the final structural design for the SCSM model.
\setlength{\tabcolsep}{8pt}
\begin{table*}[t]
	\begin{center}
		\caption{
  Ablation Study on the Loss Function Coefficients on the Loveda Dataset. The best value in each column is bolded.
		}
		\label{scsm-loss}
		\begin{tabular}{c||cccccccc}
		\Xhline{1.2pt}
            \rowcolor{mygray}
			Coefficient of $\mathcal{L}_{ce}^{d}$
&Background &Buildings &Roads &Water &Barren &Forest &Farmland&mIoU \\
			
                \hline \hline
                0.4 &47.3	&56.1	&57.8	&78.9	&18.2	&44.8	&64.6	&52.5\\
			0.8 &\bf48.3	&\bf60.4	&\bf58.4	&80.7	&\bf19.6	&47.6	&\bf67.2	&\bf54.6\\
			1.0 &48.1	&59.9	&58.1	&80.3	&19.3	&\bf47.7	&66.6	&54.3\\
			\Xhline{1.1pt}
		\end{tabular}
	\end{center}
\end{table*}

\begin{figure*}[t]
	\centering \includegraphics[width=1.0\textwidth]
       {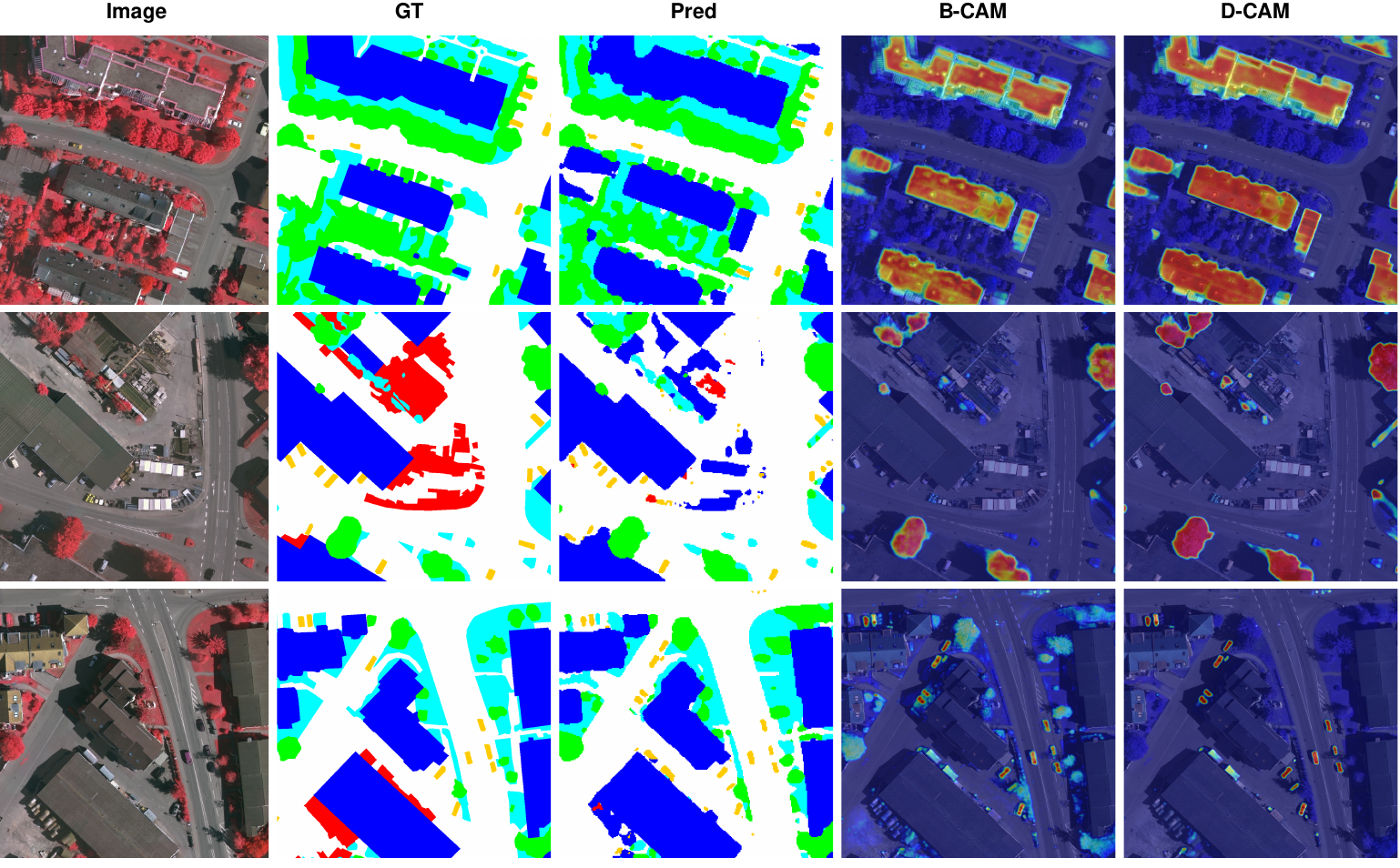}
	\centering
	\caption{Comparing the feature maps at different stages, B-CAM represents the features from the last layer of the backbone network, while D-CAM represents the features from the last layer after passing through the decoding head. The experiment is carried out on the vaihingen dataset. The target activation classes are building (first line), tree (second line) and car (third line), respectively.}
\label{fig:cam}
\end{figure*}

In addition, we provide visual evidence to further validate the effectiveness of the proposed SCA and SMG modules, as shown in Fig. \ref{fig:ablation_cam}. Specifically, we provide class activation maps for Base+GA, Base+SCA, Base+SCA+SMG (G+G), and SCSM. It can be observed that Base+GA is relatively less accurate for the activation regions of target classes (e.g., buildings and cars) and can suffer from boundary blurring. While the activation region of Base+SCA basically overlaps with the target objects, which verifies from a visual perspective that scene coupling can help improve the performance of semantic segmentation in complex scenes. In addition, Base+SCA+SMG (G+G) can further enhance the activation strength of the target region, i.e., the semantic discriminative property of the object is improved. By combining SCA and SMG, the proposed SCSM significantly improves the activation strength and accuracy in the target object, and reduces the erroneous activation regions. Quantitative and qualitative analyses demonstrate that SCSM can help to enhance the segmentation performance of geospatial objects.

\subsubsection{Ablation Study on ROPE+}
We perform an ablation analysis on the LoveDA dataset to verify the effectiveness of the ROPE+ module. As shown in Table \ref{table-rope}, it can be observed that when the ROPE+ module is removed, the model reduces the IoU for each category and the mIoU to some extent. This can be interpreted as a reduction in the model's ability to model the distribution of objects within the scene when the ROPE module is removed. Thus, the final segmentation performance is degraded. In addition, we use two widely adopted positional encoding methods, i.e., the Sinusoidal positional encoding \cite{detr} and the conditional positional encoding \cite{cpvt}, in the SCA module to further explore the effectiveness of ROPE+. As shown in Table \ref{table-rope}, when Sinusoidal positional encoding is applied, the segmentation performance of the model is lower compared to the variant without positional encoding. A plausible explanation is that Sinusoidal position encoding is more inclined to model the absolute positional information of the objects and does not model the object distribution within the scene well, which may impair the attentional affine process of SCSM. When conditional positional encoding is applied, the performance is improved compared to the variant without positional encoding, but still lower than the version with ROPE+. Extensive experimental results validate the effectiveness of ROPE+.

\subsubsection{Ablation Study on Loss Functions}
The paper conducts ablation experiments to identify the optimal combination of loss function coefficients, as shown in Table~\ref{scsm-loss}. The coefficients for the main loss function $\mathcal{L}_{ce}^{o}$ and the auxiliary loss function $\mathcal{L}_{ce}^{a}$ were fixed at 1.0 and 0.4, respectively. The coefficient for the pre-classification loss $\mathcal{L}_{ce}^{d}$ was varied among 0.4, 0.8, and 1.0 for the experiments. The results indicate that the model achieves the best segmentation performance when the coefficient of $\mathcal{L}_{ce}^{d}$ is set to 0.8.

\subsection{Analysis of SCSM working mechanism}
In order to further validate the effectiveness of SCSM, we conduct an in-depth analysis of the working mechanism of SCSM. Specifically, we use Class Activation Mapping (CAM) to visualize each layer of features of SCSM. As shown in Fig. \ref{fig:cam}, our target classes are building, tree and car from top to bottom. Specifically, for the first row of images, the response of the edges of buildings is significantly improved after the decode head, which can be attributed to the SCSM's ability to improve the semantic discrimination of the edges of buildings based on scene awareness. For the second row of images, the response region of trees is also significantly more complete, thus avoiding the mask fragmentation caused by the complex background. For the third row of images, the features before decode head have a weak response for the car category and lack the ability to accurately segment small objects. However, after our class-level context modeling, the car category can enhance its semantic discriminative ability by modeling the surrounding road scene, thus achieving accurate segmentation of small objects.
In conclusion, after the proposed class-wise context modeling, the response region of the target class is more accurate and stronger. This also verifies that the model's ability to segment different geospatial objects in complex scenes can be effectively enhanced by scene coupling and local global semantic mask strategies.

\setlength{\tabcolsep}{3pt}
\begin{table*}[t]
\small  
	\begin{center}
		\caption{Comparative Results of state-of-the-art method applied in Tieshan, Edong District, Hubei Province, China. Per-class best performance is marked in bold.}
		\label{table-isaid}
		\begin{tabular}{c||ccccccc|ccc|cc}
        \Xhline{1.2pt}
        \rowcolor{mygray}
        \multirow{2}{*}{ \fontsize{9}{15}\selectfont Method } 
        & { \fontsize{9}{15}\selectfont \makecell{Uncon. \\ sed.}} 
        & { \fontsize{9}{15}\selectfont Sand. } 
        & { \fontsize{9}{15}\selectfont \makecell{Carbo. \\ rock }} 
        & { \fontsize{9}{15}\selectfont Granite } 
        & { \fontsize{9}{15}\selectfont Diorite } 
        & { \fontsize{9}{15}\selectfont \makecell{Mafic \\dyke} } 
        & { \fontsize{9}{15}\selectfont Water } &
        OA &mIOU & F1 &{ \fontsize{9}{15}\selectfont \makecell{FLOPs \\(G)} } &{ \fontsize{9}{15}\selectfont \makecell{Params \\(M)} }
        \\
        \hline \hline
        
        UNet \cite{unet}&\makecell{\bf86.87\\ $\pm$ 0.36}& \makecell{59.69\\ $\pm$ 3.96} & \makecell{77.30\\ $\pm$ 0.28} & \makecell{7.37 \\ $\pm$ 6.97} & \makecell{80.44 \\ $\pm$ 0.37} & 0 & \makecell{48.62 \\ $\pm$ 2.88} & \makecell{81.96 \\ $\pm$ 0.31}&
        \makecell{40.82 \\ $\pm$ 0.71}&
        \makecell{51.47 \\ $\pm$ 1.16}&
        21.02&31.04\\

        UNet++ \cite{unet++}& \makecell{86.15 \\ $\pm$ 0.20} & \makecell{46.08 \\ $\pm$ 8.64} & \makecell{74.04 \\ $\pm$ 0.67} & \makecell{3.49 \\ $\pm$ 6.02} & \makecell{88.08 \\ $\pm$ 0.24} & \makecell{0} & \makecell{49.75 \\ $\pm$ 1.88} & \makecell{81.30 \\ $\pm$ 0.12}&
        \makecell{38.22 \\ $\pm$ 0.86}&
        \makecell{48.63 \\ $\pm$ 1.12}&        
        42.64&34.92\\

        Attention UNet \cite{attentionunet}& \makecell{86.40 \\ $\pm$ 0.26} & \makecell{45.97 \\ $\pm$ 8.47} & \makecell{75.03 \\ $\pm$ 0.83} & \makecell{0.25 \\ $\pm$ 0.43} & \makecell{81.02 \\ $\pm$ 0.22} & \makecell{0} & \makecell{\bf50.48 \\ $\pm$ 0.68} & \makecell{81.49 \\ $\pm$ 0.22}&
        \makecell{38.32 \\ $\pm$ 1.06}&
        \makecell{48.45 \\ $\pm$ 1.24}&  
        21.42&31.39\\

        DeepLab v3+ \cite{deeplabv3+}& \makecell{85.24 \\ $\pm$ 0.19} & \makecell{60.73 \\ $\pm$ 4.95} & \makecell{76.80 \\ $\pm$ 1.11} & \makecell{10.92 \\ $\pm$ 8.89} & 
        \makecell{79.19 \\ $\pm$ 0.28} & \makecell{5.21 \\ $\pm$ 4.53} & \makecell{42.11 \\ $\pm$ 3.36}&
        \makecell{80.66 \\ $\pm$ 0.20}&
        \makecell{40.18 \\ $\pm$ 0.80}&  
        \makecell{51.46 \\ $\pm$ 0.63}&  
        8.55&59.35\\    

        PSPNet \cite{pspnet}& \makecell{86.16 \\ $\pm$ 0.52} & \makecell{60.95 \\ $\pm$ 2.27} & \makecell{77.71 \\ $\pm$ 1.88} & \makecell{30.26 \\ $\pm$ 14.55} & 
        \makecell{80.56 \\ $\pm$ 0.09} & 0 & \makecell{47.84 \\ $\pm$ 4.11}&
        \makecell{81.68 \\ $\pm$ 0.58}&
        \makecell{42.92 \\ $\pm$ 1.26}&  
        \makecell{54.78 \\ $\pm$ 2.10}&
        19.17&48.70\\ 
        
        Bi-HRNet \cite{bihrnet}& \makecell{84.34\\ $\pm$ 0.57} &
        \makecell{58.70\\ $\pm$ 1.99} & \makecell{76.45\\ $\pm$ 0.17} & \makecell{6.26 \\ $\pm$ 6.05} & 
        \makecell{77.66 \\ $\pm$ 0.48} & \makecell{\bf9.26 \\ $\pm$ 4.07}&
        \makecell{41.62 \\ $\pm$ 1.08}&
        \makecell{79.54 \\ $\pm$ 0.47}&  
        \makecell{39.19 \\ $\pm$ 0.47}&
        \makecell{50.61 \\ $\pm$ 0.63}&  
        8.75&29.51 \\

        SwinUNet \cite{swinunet}& \makecell{86.34\\ $\pm$ 0.49} &
        \makecell{60.95\\ $\pm$ 1.59} & \makecell{74.55\\ $\pm$ 1.59} & \makecell{4.45 \\ $\pm$ 3.08} & 
        \makecell{80.86 \\ $\pm$ 0.26} & 0 &
        \makecell{48.47 \\ $\pm$ 1.24}&
        \makecell{79.31 \\ $\pm$ 0.23}&  
        \makecell{40.20 \\ $\pm$ 0.19}&
        \makecell{50.80 \\ $\pm$ 0.36}&  
        11.92&27.17 \\

        DPNet \cite{zhou2023lithological}& \makecell{86.58\\ $\pm$ 0.32} &
        \makecell{65.61\\ $\pm$ 0.40} & \makecell{78.95\\ $\pm$ 0.43} & \makecell{\bf33.54 \\ $\pm$ 5.48} & 
        \makecell{\bf81.24 \\ $\pm$ 0.20} & 0 &
        \makecell{49.90 \\ $\pm$ 1.23}&
        \makecell{82.35 \\ $\pm$ 0.02}&  
        \makecell{44.61 \\ $\pm$ 0.39}&
        \makecell{56.54 \\ $\pm$ 0.61}&  
        49.83&102.27 \\
\hline
        SCSM& \makecell{86.06\\ $\pm$ 0.63} &
        \makecell{\bf66.75\\ $\pm$ 0.36} & \makecell{\bf79.62\\ $\pm$ 0.55} &
        \makecell{29.19 \\ $\pm$ 2.88} &
        \makecell{80.75 \\ $\pm$ 0.87}&
        \makecell{9.19 \\ $\pm$ 3.86}&
        \makecell{49.58 \\ $\pm$ 0.75}&
        \makecell{\bf82.91 \\ $\pm$ 0.24}&
        \makecell{\bf45.95 \\ $\pm$ 0.56}&
        \makecell{\bf57.31 \\ $\pm$ 0.33}&
        
        6.54&30.19

        \\
			\Xhline{1.1pt}
		\end{tabular}
	\end{center}
    \vspace{4pt}
\end{table*}

\begin{figure*}[h]
	\centering \includegraphics[width=1.0\textwidth]
       {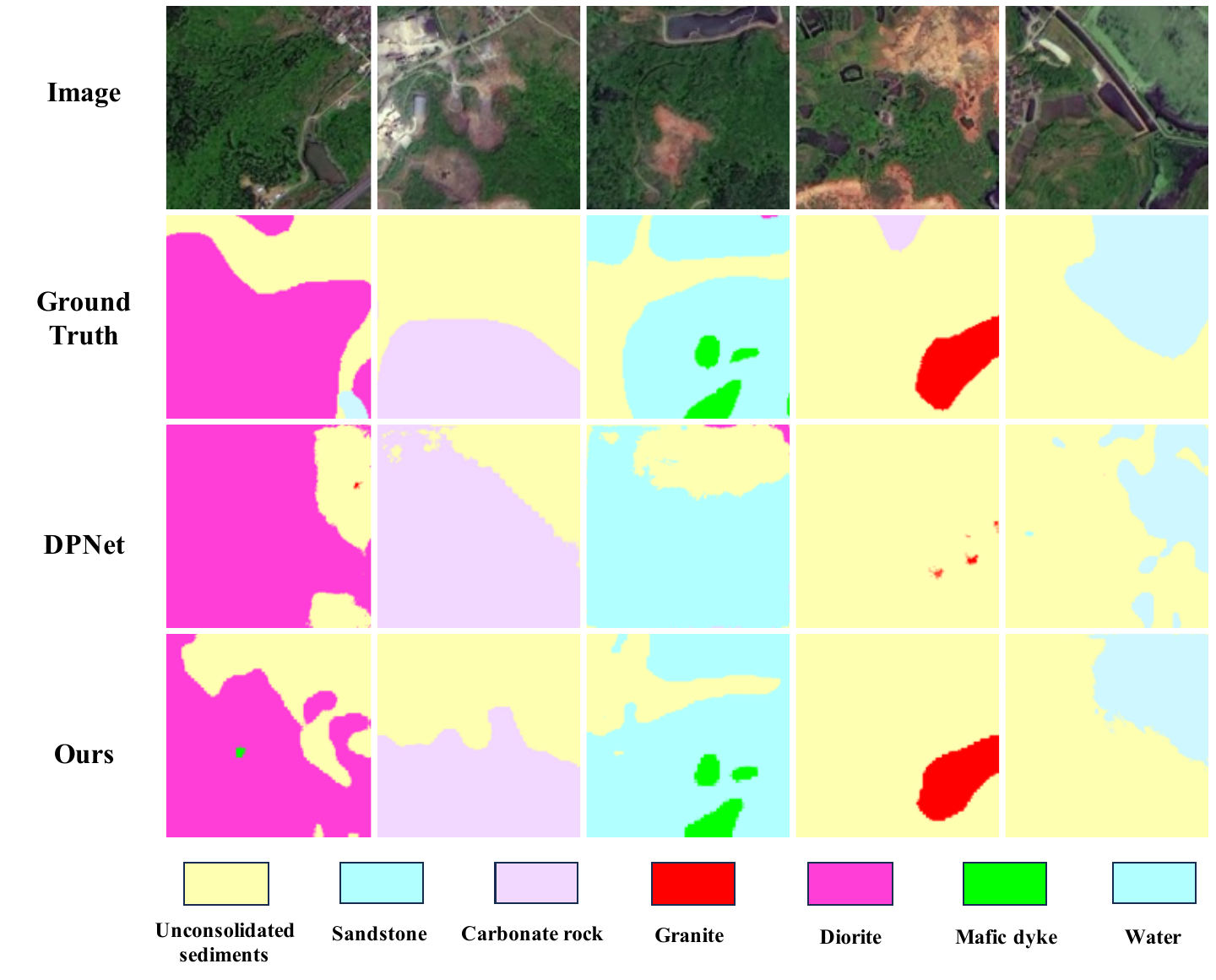}
	\centering
	\caption{Comparative visualization of state-of-the-art method applied in Tieshan, Edong District, Hubei Province, China}
\label{fig:1}
\end{figure*}

\section{Real World Exploration}
In addition to the publicly available benchmark dataset mentioned above (i.e., ISPRS Vaihingen, ISPRS Potsdam, LoveDA, iSAID), we apply SCSM to a more challenging task in the real world, i.e., Lithological Unit Classification (LUC) to better validate the effectiveness of SCSM. LUC is the classification of different types of rocks in a region, which is a sub-research area in the field of semantic segmentation of remotely sensed imagery, and has a wide range of applications in areas such as resource surveys and infrastructure planning. 
\subsection{Dataset Preparation}
Following \cite{zhou2023lithological}, our study area is in Tieshan, Edong District, Hubei Province, China, spanning 267.14 km² (coordinates: $114^\circ 45' 00''\text{E} \text{--} 115^\circ 00' 00''\text{E}$, $30^\circ 10' 00''\text{N} \text{--} 30^\circ 16' 00''\text{N}$). This region, located at the western edge of the Middle-Lower Yangtze River Metallogenic Belt, serves as a vital area for Cu, Fe, Au, and Mo extraction, hosting over 70 mineral deposits linked to Jurassic-Cretaceous intrusions \cite{lei2018ore, xie2015geochemical}.

In the 1:50,000 lithological dataset, similar units are consolidated into seven categories: unconsolidated sediments, sandstone, carbonate rock, granite, diorite, water, and mafic dykes.

The study uses two remote sensing datasets. The first, from ZiYuan-3 (2021), includes 2.1-m fused panchromatic-multispectral data and 10-m resampled DTM data. The second, from Landsat-8 (2020), provides 15-m fused multispectral data and 30-m DTM data. The 2.1-m data support remote sensing analysis, while the 30-m data, combined with 1:200,000 lithological maps, inform prior knowledge.

\subsection{Implementation Details}

The study area is divided into 45 longitudinal and 22 latitudinal regions, using a patch size of 256 × 256 pixels. Zones T and U overlap by 84.4\%, while zones 44 and 45 have a 75\% overlap.

The training, test, and validation datasets are randomly sampled from regions A–U, ensuring each dataset contains samples from all regions for spatial balance. To maintain consistent class distributions, manual adjustments are made after random sampling. The final datasets comprise 567 images for training and 189 each for testing and validation, adhering to a 6:2:2 ratio.

All experiments are repeated five times, with the mean and standard deviation calculated for comparison. The experiments are performed on a workstation equipped with eight NVIDIA Tesla V100 GPUs, each featuring 32 GB of VRAM, using PyTorch. A batch size of eight is employed, with cross-entropy serving as the loss function and RELU as the activation function.

\subsection{Results and Analysis}

We conduct a comprehensive comparison of the proposed SCSM with several state-of-the-art models, including UNet \cite{unet}, UNet++ \cite{unet++}, Attention UNet \cite{attentionunet}, DeepLab v3+ \cite{deeplabv3+}, PSPNet \cite{pspnet}, Bi-HRNet \cite{bihrnet}, SwinUNet \cite{swinunet}, and DPNet \cite{zhou2023lithological}.

The experimental results, presented in Table~\ref{table-isaid}, demonstrate that SCSM achieves state-of-the-art performance in the Tieshan area, Edong District, Hubei Province, China. Specifically, SCSM delivers improvements of 0.56\% in OA, 1.34\% in mIOU, and 0.77\% in F1 score over the recent DPNet method. Moreover, SCSM offers significant computational efficiency. It utilizes only 30.19M parameters and 6.54G FLOPs, representing a mere 29.52\% of the parameter count and 13.12\% of the computational cost required by DPNet. These improvements make SCSM particularly attractive for real-world applications where resource constraints are a concern. When compared with other advanced models such as Bi-HRNet and SwinUNet, SCSM delivers even more pronounced improvements in accuracy and achieves a better trade-off between performance and computational efficiency, setting a new benchmark for lightweight yet effective segmentation models.

To provide additional insights, we visualize segmentation maps to compare SCSM with DPNet, as shown in Fig. \ref{fig:1}. The visual results clearly indicate that SCSM is far more adept at accurately identifying object categories in complex scenarios. It excels in delineating the boundaries of object regions, achieving significantly sharper and more precise segmentation compared to DPNet. This qualitative evidence further supports the effectiveness of SCSM and its ability to handle intricate segmentation challenges with both precision and efficiency.

\section{Conclusion}
In this paper, we first examine the properties of remotely sensed imagery, including complex backgrounds, high intra-class variance and the presence of intrinsic spatial correlations between geospatial objects. These properties result in vanilla attention with limited performance due to dense affinity and lack of geospatial object perception of the scene. Based on this, we introduce two strategies, scene coupling and local global semantic masking, to reconstruct vanilla attention. The scene coupling strategy decomposes scene information into a global representation of the scene and a distribution of scene objects to be embedded in the attentional affinity process, thus effectively exploiting the intrinsic spatial correlation among geospatial objects to improve the attention modeling process. The local global semantic mask strategy uses local semantic masks with spatial prior as intermediate perceptual elements to indirectly correlate pixels with global semantic masks, which avoids foreground features from being obfuscated by a large number of background contexts and mitigates intra-class variance. In particular, we elegantly combine the two by proposing the model SCSM. SCSM possesses a highly concise formulaic representation similar to vanilla attention, which can serve as a new baseline for the exploration of attention methods in the field of remote sensing imagery. 

Extensive experiments on four benchmark datasets validate the effectiveness of SCSM. In addition, we further validate that SCSM can effectively segment various geospatial objects even in more complex real-world scenarios. This study provides valuable guidance for the direction of subsequent optimization of attention in the remote sensing community. In brief, the segmentation performance of the attention model in remote sensing images can be significantly improved by appropriately modifying the input and affinity process of the vanilla attention. In future work, we will explore more instantiation techniques along the new baseline formulation of scene-coupled semantic masks, such as employing adaptive frequency filtering to extract global representations of scenes, or optimizing the acquisition of semantic masks. We believe this can further enhance the performance of existing attention-based models in the remote sensing community.

\section{Acknowledgements}
This work is supported by the Public Welfare Science and Technology Plan of Ningbo City (2022S125) and the Key Research and Development Plan of Zhejiang Province (2021C01031). In addition, we are particularly thankful to \cite{zhou2023lithological} for providing the LUC dataset in Tieshan, Edong District, Hubei Province, China.
\bibliographystyle{elsarticle-num}

\bibliography{cas-refs}

\begin{thebibliography}{100}
\expandafter\ifx\csname url\endcsname\relax
  \def\url#1{\texttt{#1}}\fi
\expandafter\ifx\csname urlprefix\endcsname\relax\def\urlprefix{URL }\fi
\expandafter\ifx\csname href\endcsname\relax
  \def\href#1#2{#2} \def\path#1{#1}\fi

\bibitem{urban}
Q.~Zhang, K.~C. Seto, Mapping urbanization dynamics at regional and global scales using multi-temporal dmsp/ols nighttime light data, Remote Sensing of Environment 115~(9) (2011) 2320--2329.

\bibitem{urban2}
B.~Huang, B.~Zhao, Y.~Song, Urban land-use mapping using a deep convolutional neural network with high spatial resolution multispectral remote sensing imagery, Remote Sensing of Environment 214 (2018) 73--86.

\bibitem{he2024geolocation}
J.~He, T.~Nie, W.~Ma, Geolocation representation from large language models are generic enhancers for spatio-temporal learning, arXiv preprint arXiv:2408.12116 (2024).

\bibitem{environment}
Q.~Yuan, H.~Shen, T.~Li, Z.~Li, S.~Li, Y.~Jiang, H.~Xu, W.~Tan, Q.~Yang, J.~Wang, et~al., Deep learning in environmental remote sensing: Achievements and challenges, Remote Sensing of Environment 241 (2020) 111716.

\bibitem{cui2024real}
K.~Cui, W.~Tang, R.~Zhu, M.~Wang, G.~D. Larsen, V.~P. Pauca, S.~Alqahtani, F.~Yang, D.~Segurado, P.~Fine, et~al., Real-time localization and bimodal point pattern analysis of palms using uav imagery, arXiv preprint arXiv:2410.11124 (2024).

\bibitem{road}
M.~Maboudi, J.~Amini, S.~Malihi, M.~Hahn, Integrating fuzzy object based image analysis and ant colony optimization for road extraction from remotely sensed images, ISPRS Journal of Photogrammetry and Remote Sensing 138 (2018) 151--163.

\bibitem{wang2024airshot}
Z.~Wang, B.~Li, C.~Wang, S.~Scherer, \href{https://arxiv.org/pdf/2404.05069.pdf}{{AirShot}: Efficient few-shot detection for autonomous exploration}, in: IEEE/RSJ International Conference on Intelligent Robots and Systems (IROS), 2024.
\newline\urlprefix\url{https://arxiv.org/pdf/2404.05069.pdf}

\bibitem{wang2024onls}
Z.~Wang, \href{https://openreview.net/forum?id=Q8diCUHTZd}{{ONLS}: {OPTIMAL} {NOISE} {LEVEL} {SEARCH} {IN} {DIFFUSION} {AUTOENCODERS} {WITHOUT} {FINE}-{TUNING}}, in: The Second Tiny Papers Track at ICLR 2024, 2024.
\newline\urlprefix\url{https://openreview.net/forum?id=Q8diCUHTZd}

\bibitem{rsdataset}
F.~Bastani, P.~Wolters, R.~Gupta, J.~Ferdinando, A.~Kembhavi, Satlaspretrain: A large-scale dataset for remote sensing image understanding, in: Proceedings of the IEEE/CVF International Conference on Computer Vision (ICCV), 2023, pp. 16772--16782.

\bibitem{liu2024crossmatch}
R.~Liu, T.~Luo, S.~Huang, Y.~Wu, Z.~Jiang, H.~Zhang, Crossmatch: Cross-view matching for semi-supervised remote sensing image segmentation, IEEE Transactions on Geoscience and Remote Sensing (2024).

\bibitem{10647298}
T.~Luo, M.~Du, J.~Shi, X.~Chen, B.~Zhao, S.~Huang, Contextuality helps representation learning for generalized category discovery, in: 2024 IEEE International Conference on Image Processing (ICIP), 2024, pp. 687--693.
\newblock \href {https://doi.org/10.1109/ICIP51287.2024.10647298} {\path{doi:10.1109/ICIP51287.2024.10647298}}.

\bibitem{cui2024superpixel}
K.~Cui, R.~Li, S.~L. Polk, Y.~Lin, H.~Zhang, J.~M. Murphy, R.~J. Plemmons, R.~H. Chan, Superpixel-based and spatially-regularized diffusion learning for unsupervised hyperspectral image clustering, IEEE Transactions on Geoscience and Remote Sensing (2024).

\bibitem{cui2022unsupervised}
K.~Cui, R.~Li, S.~L. Polk, J.~M. Murphy, R.~J. Plemmons, R.~H. Chan, Unsupervised spatial-spectral hyperspectral image reconstruction and clustering with diffusion geometry, in: 2022 12th Workshop on Hyperspectral Imaging and Signal Processing: Evolution in Remote Sensing (WHISPERS), IEEE, 2022, pp. 1--5.

\bibitem{chen2024bimcv}
Y.~Chen, C.~Liu, X.~Liu, R.~Arcucci, Z.~Xiong, Bimcv-r: A landmark dataset for 3d ct text-image retrieval, in: MICCAI, Springer, 2024, pp. 124--134.

\bibitem{chen2024learning}
Y.~Chen, W.~Huang, X.~Liu, S.~Deng, Q.~Chen, Z.~Xiong, Learning multiscale consistency for self-supervised electron microscopy instance segmentation, in: ICASSP, IEEE, 2024, pp. 1566--1570.

\bibitem{significant}
J.~Xue, B.~Su, Significant remote sensing vegetation indices: A review of developments and applications, Journal of sensors 2017~(1) (2017) 1353691.

\bibitem{fcn}
J.~Long, E.~Shelhamer, T.~Darrell, Fully convolutional networks for semantic segmentation, in: Proceedings of the IEEE conference on computer vision and pattern recognition, 2015, pp. 3431--3440.

\bibitem{transformer}
A.~Vaswani, N.~Shazeer, N.~Parmar, J.~Uszkoreit, L.~Jones, A.~N. Gomez, {\L}.~Kaiser, I.~Polosukhin, Attention is all you need, Advances in neural information processing systems 30 (2017).

\bibitem{zhang2023body}
Y.~Zhang, P.~Ji, A.~Wang, J.~Mei, A.~Kortylewski, A.~L. Yuille, \href{https://doi.org/10.1109/ICCV51070.2023.00862}{3d-aware neural body fitting for occlusion robust 3d human pose estimation}, in: IEEE/CVF International Conference on Computer Vision (ICCV), 2023, pp. 9365--9376.
\newblock \href {https://doi.org/10.1109/ICCV51070.2023.00862} {\path{doi:10.1109/ICCV51070.2023.00862}}.
\newline\urlprefix\url{https://doi.org/10.1109/ICCV51070.2023.00862}

\bibitem{wang2024sparse}
Y.~Wang, Y.~Zhang, M.~Huo, R.~Tian, X.~Zhang, Y.~Xie, C.~Xu, P.~Ji, W.~Zhan, M.~Ding, M.~Tomizuka, \href{https://doi.org/10.48550/arXiv.2407.01531}{Sparse diffusion policy: A sparse, reusable, and flexible policy for robot learning}, CoRR abs/2407.01531 (2024).
\newblock \href {https://doi.org/10.48550/ARXIV.2407.01531} {\path{doi:10.48550/ARXIV.2407.01531}}.
\newline\urlprefix\url{https://doi.org/10.48550/arXiv.2407.01531}

\bibitem{nie2024imputeformer}
T.~Nie, G.~Qin, W.~Ma, Y.~Mei, J.~Sun, Imputeformer: Low rankness-induced transformers for generalizable spatiotemporal imputation, in: Proceedings of the 30th ACM SIGKDD Conference on Knowledge Discovery and Data Mining, 2024, pp. 2260--2271.

\bibitem{trias2008using}
R.~Trias-Sanz, G.~Stamon, J.~Louchet, Using colour, texture, and hierarchial segmentation for high-resolution remote sensing, ISPRS Journal of Photogrammetry and remote sensing 63~(2) (2008) 156--168.

\bibitem{jin2022imc}
J.~Jin, H.~Xu, P.~Ji, B.~Leng, \href{https://doi.org/10.1109/ICIP46576.2022.9897709}{Imc-net: Learning implicit field with corner attention network for 3d shape reconstruction}, in: IEEE International Conference on Image Processing (ICIP), 2022, pp. 1591--1595.
\newblock \href {https://doi.org/10.1109/ICIP46576.2022.9897709} {\path{doi:10.1109/ICIP46576.2022.9897709}}.
\newline\urlprefix\url{https://doi.org/10.1109/ICIP46576.2022.9897709}

\bibitem{bi2024decoding}
B.~Bi, S.~Liu, L.~Mei, Y.~Wang, P.~Ji, X.~Cheng, \href{https://doi.org/10.48550/arXiv.2405.11613}{Decoding by contrasting knowledge: Enhancing llms' confidence on edited facts}, CoRR abs/2405.11613 (2024).
\newblock \href {https://doi.org/10.48550/ARXIV.2405.11613} {\path{doi:10.48550/ARXIV.2405.11613}}.
\newline\urlprefix\url{https://doi.org/10.48550/arXiv.2405.11613}

\bibitem{chen2023self}
Y.~Chen, W.~Huang, S.~Zhou, Q.~Chen, Z.~Xiong, Self-supervised neuron segmentation with multi-agent reinforcement learning, in: IJCAI, 2023, pp. 609--617.

\bibitem{qianmaskfactory}
H.~Qian, Y.~Chen, S.~Lou, F.~Khan, X.~Jin, D.-P. Fan, Maskfactory: Towards high-quality synthetic data generation for dichotomous image segmentation, in: NeurIPS, 2024.

\bibitem{pspnet}
H.~Zhao, J.~Shi, X.~Qi, X.~Wang, J.~Jia, Pyramid scene parsing network, in: Proceedings of the IEEE conference on computer vision and pattern recognition, 2017, pp. 2881--2890.

\bibitem{deeplabv3+}
L.-C. Chen, Y.~Zhu, G.~Papandreou, F.~Schroff, H.~Adam, Encoder-decoder with atrous separable convolution for semantic image segmentation, in: Proceedings of the European conference on computer vision (ECCV), 2018, pp. 801--818.

\bibitem{ocrnet}
Y.~Yuan, X.~Chen, J.~Wang, Object-contextual representations for semantic segmentation, in: Computer Vision--ECCV 2020: 16th European Conference, Glasgow, UK, August 23--28, 2020, Proceedings, Part VI 16, Springer, 2020, pp. 173--190.

\bibitem{chen2024tokenunify}
Y.~Chen, H.~Shi, X.~Liu, T.~Shi, R.~Zhang, D.~Liu, Z.~Xiong, F.~Wu, Tokenunify: Scalable autoregressive visual pre-training with mixture token prediction, arXiv preprint arXiv:2405.16847 (2024).

\bibitem{dmnet}
J.~He, Z.~Deng, Y.~Qiao, Dynamic multi-scale filters for semantic segmentation, in: Proceedings of the IEEE/CVF International Conference on Computer Vision, 2019, pp. 3562--3572.

\bibitem{isnet}
Z.~Jin, B.~Liu, Q.~Chu, N.~Yu, Isnet: Integrate image-level and semantic-level context for semantic segmentation, in: Proceedings of the IEEE/CVF International Conference on Computer Vision, 2021, pp. 7189--7198.

\bibitem{logcan++}
X.~Ma, R.~Lian, Z.~Wu, H.~Guo, M.~Ma, S.~Wu, Z.~Du, S.~Song, W.~Zhang, \href{https://arxiv.org/abs/2406.16502}{Logcan++: Adaptive local-global class-aware network for semantic segmentation of remote sensing imagery} (2024).
\newblock \href {http://arxiv.org/abs/2406.16502} {\path{arXiv:2406.16502}}.
\newline\urlprefix\url{https://arxiv.org/abs/2406.16502}

\bibitem{danet}
J.~Fu, J.~Liu, H.~Tian, Y.~Li, Y.~Bao, Z.~Fang, H.~Lu, Dual attention network for scene segmentation, in: Proceedings of the IEEE/CVF conference on computer vision and pattern recognition, 2019, pp. 3146--3154.

\bibitem{senet}
J.~Hu, L.~Shen, G.~Sun, Squeeze-and-excitation networks, in: Proceedings of the IEEE conference on computer vision and pattern recognition, 2018, pp. 7132--7141.

\bibitem{docnet}
X.~Ma, R.~Che, X.~Wang, M.~Ma, S.~Wu, T.~Feng, W.~Zhang, Docnet: Dual-domain optimized class-aware network for remote sensing image segmentation, IEEE Geoscience and Remote Sensing Letters (2024).

\bibitem{ccnet}
Z.~Huang, X.~Wang, L.~Huang, C.~Huang, Y.~Wei, W.~Liu, Ccnet: Criss-cross attention for semantic segmentation, in: Proceedings of the IEEE/CVF international conference on computer vision, 2019, pp. 603--612.

\bibitem{flanet}
Q.~Song, J.~Li, C.~Li, H.~Guo, R.~Huang, Fully attentional network for semantic segmentation, in: Proceedings of the AAAI Conference on Artificial Intelligence, Vol.~36, 2022, pp. 2280--2288.

\bibitem{pointflow}
X.~Li, H.~He, X.~Li, D.~Li, G.~Cheng, J.~Shi, L.~Weng, Y.~Tong, Z.~Lin, Pointflow: Flowing semantics through points for aerial image segmentation, in: Proceedings of the IEEE/CVF Conference on Computer Vision and Pattern Recognition, 2021, pp. 4217--4226.

\bibitem{mamba}
A.~Gu, T.~Dao, Mamba: Linear-time sequence modeling with selective state spaces, arXiv preprint arXiv:2312.00752 (2023).

\bibitem{mamba2}
T.~Dao, A.~Gu, Transformers are ssms: Generalized models and efficient algorithms through structured state space duality, arXiv preprint arXiv:2405.21060 (2024).

\bibitem{swinunet}
X.~He, Y.~Zhou, J.~Zhao, D.~Zhang, R.~Yao, Y.~Xue, Swin transformer embedding unet for remote sensing image semantic segmentation, IEEE Transactions on Geoscience and Remote Sensing 60 (2022) 1--15.

\bibitem{zhang2022transformer}
C.~Zhang, W.~Jiang, Y.~Zhang, W.~Wang, Q.~Zhao, C.~Wang, Transformer and cnn hybrid deep neural network for semantic segmentation of very-high-resolution remote sensing imagery, IEEE Transactions on Geoscience and Remote Sensing 60 (2022) 1--20.

\bibitem{ding2022looking}
L.~Ding, D.~Lin, S.~Lin, J.~Zhang, X.~Cui, Y.~Wang, H.~Tang, L.~Bruzzone, Looking outside the window: Wide-context transformer for the semantic segmentation of high-resolution remote sensing images, IEEE Transactions on Geoscience and Remote Sensing 60 (2022) 1--13.

\bibitem{glots}
Y.~Liu, Y.~Zhang, Y.~Wang, S.~Mei, Rethinking transformers for semantic segmentation of remote sensing images, IEEE Transactions on Geoscience and Remote Sensing (2023).

\bibitem{sun2024ultra}
H.~Sun, \href{https://arxiv.org/abs/2412.10181}{Ultra-high resolution segmentation via boundary-enhanced patch-merging transformer} (2024).
\newblock \href {http://arxiv.org/abs/2412.10181} {\path{arXiv:2412.10181}}.
\newline\urlprefix\url{https://arxiv.org/abs/2412.10181}

\bibitem{cheng2024spt}
S.~Cheng, H.~Sun, \href{https://arxiv.org/abs/2412.10224}{Spt: Sequence prompt transformer for interactive image segmentation} (2024).
\newblock \href {http://arxiv.org/abs/2412.10224} {\path{arXiv:2412.10224}}.
\newline\urlprefix\url{https://arxiv.org/abs/2412.10224}

\bibitem{rsmamba}
S.~Zhao, H.~Chen, X.~Zhang, P.~Xiao, L.~Bai, W.~Ouyang, Rs-mamba for large remote sensing image dense prediction, arXiv preprint arXiv:2404.02668 (2024).

\bibitem{resnet}
K.~He, X.~Zhang, S.~Ren, J.~Sun, Deep residual learning for image recognition, in: Proceedings of the IEEE conference on computer vision and pattern recognition, 2016, pp. 770--778.

\bibitem{denseaspp}
M.~Yang, K.~Yu, C.~Zhang, Z.~Li, K.~Yang, Denseaspp for semantic segmentation in street scenes, in: Proceedings of the IEEE conference on computer vision and pattern recognition, 2018, pp. 3684--3692.

\bibitem{caa}
Y.~Huang, D.~Kang, W.~Jia, L.~Liu, X.~He, Channelized axial attention--considering channel relation within spatial attention for semantic segmentation, in: Proceedings of the AAAI Conference on Artificial Intelligence, Vol.~36, 2022, pp. 1016--1025.

\bibitem{ccanet}
G.~Deng, Z.~Wu, C.~Wang, M.~Xu, Y.~Zhong, Ccanet: Class-constraint coarse-to-fine attentional deep network for subdecimeter aerial image semantic segmentation, IEEE Transactions on Geoscience and Remote Sensing 60 (2022) 1--20.
\newblock \href {https://doi.org/10.1109/TGRS.2021.3055950} {\path{doi:10.1109/TGRS.2021.3055950}}.

\bibitem{gmmseg}
C.~Liang, W.~Wang, J.~Miao, Y.~Yang, Gmmseg: Gaussian mixture based generative semantic segmentation models, arXiv preprint arXiv:2210.02025 (2022).

\bibitem{protoseg}
T.~Zhou, W.~Wang, E.~Konukoglu, L.~Van~Gool, Rethinking semantic segmentation: A prototype view, in: Proceedings of the IEEE/CVF Conference on Computer Vision and Pattern Recognition, 2022, pp. 2582--2593.

\bibitem{vit}
A.~Dosovitskiy, L.~Beyer, A.~Kolesnikov, D.~Weissenborn, X.~Zhai, T.~Unterthiner, M.~Dehghani, M.~Minderer, G.~Heigold, S.~Gelly, et~al., An image is worth 16x16 words: Transformers for image recognition at scale, arXiv preprint arXiv:2010.11929 (2020).

\bibitem{segformer}
E.~Xie, W.~Wang, Z.~Yu, A.~Anandkumar, J.~M. Alvarez, P.~Luo, Segformer: Simple and efficient design for semantic segmentation with transformers, Advances in Neural Information Processing Systems 34 (2021) 12077--12090.

\bibitem{swintransformer}
Z.~Liu, Y.~Lin, Y.~Cao, H.~Hu, Y.~Wei, Z.~Zhang, S.~Lin, B.~Guo, Swin transformer: Hierarchical vision transformer using shifted windows, in: Proceedings of the IEEE/CVF international conference on computer vision, 2021, pp. 10012--10022.

\bibitem{mlp}
I.~O. Tolstikhin, N.~Houlsby, A.~Kolesnikov, L.~Beyer, X.~Zhai, T.~Unterthiner, J.~Yung, A.~Steiner, D.~Keysers, J.~Uszkoreit, et~al., Mlp-mixer: An all-mlp architecture for vision, Advances in neural information processing systems 34 (2021) 24261--24272.

\bibitem{mlps}
H.~Liu, Z.~Dai, D.~So, Q.~V. Le, Pay attention to mlps, Advances in Neural Information Processing Systems 34 (2021) 9204--9215.

\bibitem{maskformer}
B.~Cheng, A.~Schwing, A.~Kirillov, Per-pixel classification is not all you need for semantic segmentation, Advances in Neural Information Processing Systems 34 (2021) 17864--17875.

\bibitem{wu2024domain}
Z.~Wu, J.~Lu, J.~Han, L.~Bai, Y.~Zhang, Z.~Zhao, S.~Song, Domain separation graph neural networks for saliency object ranking, in: Proceedings of the IEEE/CVF Conference on Computer Vision and Pattern Recognition, 2024, pp. 3964--3974.

\bibitem{beit}
H.~Bao, L.~Dong, S.~Piao, F.~Wei, Beit: Bert pre-training of image transformers, arXiv preprint arXiv:2106.08254 (2021).

\bibitem{sunprogram}
H.~Sun, L.~Xu, S.~Jin, P.~Luo, C.~Qian, W.~Liu, Program: Prototype graph model based pseudo-label learning for test-time adaptation, in: The Twelfth International Conference on Learning Representations, 2024.

\bibitem{dilateformer}
J.~Jiao, Y.-M. Tang, K.-Y. Lin, Y.~Gao, J.~Ma, Y.~Wang, W.-S. Zheng, Dilateformer: Multi-scale dilated transformer for visual recognition, IEEE Transactions on Multimedia (2023).

\bibitem{segmenter}
R.~Strudel, R.~Garcia, I.~Laptev, C.~Schmid, Segmenter: Transformer for semantic segmentation, in: Proceedings of the IEEE/CVF international conference on computer vision, 2021, pp. 7262--7272.

\bibitem{road1}
L.~Dai, G.~Zhang, R.~Zhang, Radanet: Road augmented deformable attention network for road extraction from complex high-resolution remote-sensing images, IEEE Transactions on Geoscience and Remote Sensing (2023).

\bibitem{bdtnet}
L.~Luo, J.-X. Wang, S.-B. Chen, J.~Tang, B.~Luo, Bdtnet: Road extraction by bi-direction transformer from remote sensing images, IEEE Geoscience and Remote Sensing Letters 19 (2022) 1--5.

\bibitem{bmda}
S.~Dong, Z.~Chen, Block multi-dimensional attention for road segmentation in remote sensing imagery, IEEE Geoscience and Remote Sensing Letters 19 (2021) 1--5.

\bibitem{building1}
H.~Jung, H.-S. Choi, M.~Kang, Boundary enhancement semantic segmentation for building extraction from remote sensed image, IEEE Transactions on Geoscience and Remote Sensing 60 (2021) 1--12.

\bibitem{lbe}
J.~Yuan, Learning building extraction in aerial scenes with convolutional networks, IEEE transactions on pattern analysis and machine intelligence 40~(11) (2017) 2793--2798.

\bibitem{land2}
A.~Alem, S.~Kumar, Transfer learning models for land cover and land use classification in remote sensing image, Applied Artificial Intelligence 36~(1) (2022) 2014192.

\bibitem{jdp}
C.~Zhang, I.~Sargent, X.~Pan, H.~Li, A.~Gardiner, J.~Hare, P.~M. Atkinson, Joint deep learning for land cover and land use classification, Remote sensing of environment 221 (2019) 173--187.

\bibitem{mdanet}
R.~Zuo, G.~Zhang, R.~Zhang, X.~Jia, A deformable attention network for high-resolution remote sensing images semantic segmentation, IEEE Transactions on Geoscience and Remote Sensing 60 (2022) 1--14.
\newblock \href {https://doi.org/10.1109/TGRS.2021.3119537} {\path{doi:10.1109/TGRS.2021.3119537}}.

\bibitem{SSGCC}
R.~Guan, W.~Tu, Z.~Li, H.~Yu, D.~Hu, Y.~Chen, C.~Tang, Q.~Yuan, X.~Liu, Spatial-spectral graph contrastive clustering with hard sample mining for hyperspectral images, IEEE Transactions on Geoscience and Remote Sensing (2024) 1--16.

\bibitem{CMSCGC}
R.~Guan, Z.~Li, W.~Tu, J.~Wang, Y.~Liu, X.~Li, C.~Tang, R.~Feng, Contrastive multiview subspace clustering of hyperspectral images based on graph convolutional networks, IEEE Transactions on Geoscience and Remote Sensing 62 (2024) 1--14.

\bibitem{manet}
R.~Li, S.~Zheng, C.~Zhang, C.~Duan, J.~Su, L.~Wang, P.~M. Atkinson, Multiattention network for semantic segmentation of fine-resolution remote sensing images, IEEE Transactions on Geoscience and Remote Sensing 60 (2021) 1--13.

\bibitem{farseg}
Z.~Zheng, Y.~Zhong, J.~Wang, A.~Ma, Foreground-aware relation network for geospatial object segmentation in high spatial resolution remote sensing imagery, in: Proceedings of the IEEE/CVF conference on computer vision and pattern recognition, 2020, pp. 4096--4105.

\bibitem{sco}
F.~Yang, C.~Ma, Sparse and complete latent organization for geospatial semantic segmentation, in: Proceedings of the IEEE/CVF Conference on Computer Vision and Pattern Recognition, 2022, pp. 1809--1818.

\bibitem{nonlocal}
X.~Wang, R.~Girshick, A.~Gupta, K.~He, Non-local neural networks, in: Proceedings of the IEEE conference on computer vision and pattern recognition, 2018, pp. 7794--7803.

\bibitem{vaihingen}
F.~Rottensteiner, G.~Sohn, J.~Jung, M.~Gerke, C.~Baillard, S.~Bnitez, U.~Breitkopf, International society for photogrammetry and remote sensing, 2d semantic labeling contest, available: \url{https://www.isprs.org/education/benchmarks/UrbanSemLab} (Accessed: Oct. 29, 2020.).

\bibitem{loveda}
J.~Wang, Z.~Zheng, A.~Ma, X.~Lu, Y.~Zhong, Loveda: A remote sensing land-cover dataset for domain adaptive semantic segmentation, arXiv preprint arXiv:2110.08733 (2021).

\bibitem{isaid}
S.~Waqas~Zamir, A.~Arora, A.~Gupta, S.~Khan, G.~Sun, F.~Shahbaz~Khan, F.~Zhu, L.~Shao, G.-S. Xia, X.~Bai, isaid: A large-scale dataset for instance segmentation in aerial images, in: Proceedings of the IEEE/CVF Conference on Computer Vision and Pattern Recognition Workshops, 2019, pp. 28--37.

\bibitem{farseg++}
Z.~Zheng, Y.~Zhong, J.~Wang, A.~Ma, L.~Zhang, Farseg++: Foreground-aware relation network for geospatial object segmentation in high spatial resolution remote sensing imagery, IEEE Transactions on Pattern Analysis and Machine Intelligence (2023).

\bibitem{rope}
J.~Su, M.~Ahmed, Y.~Lu, S.~Pan, W.~Bo, Y.~Liu, Roformer: Enhanced transformer with rotary position embedding, Neurocomputing 568 (2024) 127063.

\bibitem{relative}
P.~Shaw, J.~Uszkoreit, A.~Vaswani, Self-attention with relative position representations, arXiv preprint arXiv:1803.02155 (2018).

\bibitem{fcanet}
Z.~Qin, P.~Zhang, F.~Wu, X.~Li, Fcanet: Frequency channel attention networks, in: Proceedings of the IEEE/CVF international conference on computer vision, 2021, pp. 783--792.

\bibitem{aff}
Z.~Huang, Z.~Zhang, C.~Lan, Z.-J. Zha, Y.~Lu, B.~Guo, Adaptive frequency filters as efficient global token mixers, in: Proceedings of the IEEE/CVF International Conference on Computer Vision, 2023, pp. 6049--6059.

\bibitem{fpn}
A.~Kirillov, R.~Girshick, K.~He, P.~Doll{\'a}r, Panoptic feature pyramid networks, in: Proceedings of the IEEE/CVF conference on computer vision and pattern recognition, 2019, pp. 6399--6408.

\bibitem{lanet}
L.~Ding, H.~Tang, L.~Bruzzone, Lanet: Local attention embedding to improve the semantic segmentation of remote sensing images, IEEE Transactions on Geoscience and Remote Sensing 59~(1) (2020) 426--435.

\bibitem{convnext}
Z.~Liu, H.~Mao, C.-Y. Wu, C.~Feichtenhofer, T.~Darrell, S.~Xie, A convnet for the 2020s, in: Proceedings of the IEEE/CVF Conference on Computer Vision and Pattern Recognition, 2022, pp. 11976--11986.

\bibitem{poolformer}
W.~Yu, M.~Luo, P.~Zhou, C.~Si, Y.~Zhou, X.~Wang, J.~Feng, S.~Yan, Metaformer is actually what you need for vision, in: Proceedings of the IEEE/CVF conference on computer vision and pattern recognition, 2022, pp. 10819--10829.

\bibitem{biformer}
L.~Zhu, X.~Wang, Z.~Ke, W.~Zhang, R.~W. Lau, Biformer: Vision transformer with bi-level routing attention, in: Proceedings of the IEEE/CVF Conference on Computer Vision and Pattern Recognition, 2023, pp. 10323--10333.

\bibitem{efficientvit}
H.~Cai, J.~Li, M.~Hu, C.~Gan, S.~Han, Efficientvit: Lightweight multi-scale attention for high-resolution dense prediction, in: Proceedings of the IEEE/CVF International Conference on Computer Vision, 2023, pp. 17302--17313.

\bibitem{ddp}
Y.~Ji, Z.~Chen, E.~Xie, L.~Hong, X.~Liu, Z.~Liu, T.~Lu, Z.~Li, P.~Luo, Ddp: Diffusion model for dense visual prediction, in: Proceedings of the IEEE/CVF International Conference on Computer Vision, 2023, pp. 21741--21752.

\bibitem{imagenet}
J.~Deng, W.~Dong, R.~Socher, L.-J. Li, K.~Li, L.~Fei-Fei, Imagenet: A large-scale hierarchical image database, in: 2009 IEEE conference on computer vision and pattern recognition, Ieee, 2009, pp. 248--255.

\bibitem{acfnet}
F.~Zhang, Y.~Chen, Z.~Li, Z.~Hong, J.~Liu, F.~Ma, J.~Han, E.~Ding, Acfnet: Attentional class feature network for semantic segmentation, in: Proceedings of the IEEE/CVF International Conference on Computer Vision, 2019, pp. 6798--6807.

\bibitem{spgnet}
B.~Cheng, L.-C. Chen, Y.~Wei, Y.~Zhu, Z.~Huang, J.~Xiong, T.~S. Huang, W.-M. Hwu, H.~Shi, Spgnet: Semantic prediction guidance for scene parsing, in: Proceedings of the IEEE/CVF international conference on computer vision, 2019, pp. 5218--5228.

\bibitem{deeplabv3}
L.-C. Chen, G.~Papandreou, F.~Schroff, H.~Adam, Rethinking atrous convolution for semantic image segmentation, arXiv preprint arXiv:1706.05587 (2017).

\bibitem{unet}
O.~Ronneberger, P.~Fischer, T.~Brox, U-net: Convolutional networks for biomedical image segmentation, in: Medical image computing and computer-assisted intervention--MICCAI 2015: 18th international conference, Munich, Germany, October 5-9, 2015, proceedings, part III 18, Springer, 2015, pp. 234--241.

\bibitem{dnlnet}
M.~Yin, Z.~Yao, Y.~Cao, X.~Li, Z.~Zhang, S.~Lin, H.~Hu, Disentangled non-local neural networks, in: Computer Vision--ECCV 2020: 16th European Conference, Glasgow, UK, August 23--28, 2020, Proceedings, Part XV 16, Springer, 2020, pp. 191--207.

\bibitem{gcnet}
Y.~Cao, J.~Xu, S.~Lin, F.~Wei, H.~Hu, Global context networks, IEEE Transactions on Pattern Analysis and Machine Intelligence 45~(6) (2020) 6881--6895.

\bibitem{ocnet}
Y.~Yuan, X.~Chen, J.~Wang, Object-contextual representations for semantic segmentation, in: Computer Vision--ECCV 2020: 16th European Conference, Glasgow, UK, August 23--28, 2020, Proceedings, Part VI 16, Springer, 2020, pp. 173--190.

\bibitem{emanet}
X.~Li, Z.~Zhong, J.~Wu, Y.~Yang, Z.~Lin, H.~Liu, Expectation-maximization attention networks for semantic segmentation, in: Proceedings of the IEEE/CVF international conference on computer vision, 2019, pp. 9167--9176.

\bibitem{hrnet}
J.~Wang, K.~Sun, T.~Cheng, B.~Jiang, C.~Deng, Y.~Zhao, D.~Liu, Y.~Mu, M.~Tan, X.~Wang, et~al., Deep high-resolution representation learning for visual recognition, IEEE transactions on pattern analysis and machine intelligence 43~(10) (2020) 3349--3364.

\bibitem{upernet}
T.~Xiao, Y.~Liu, B.~Zhou, Y.~Jiang, J.~Sun, Unified perceptual parsing for scene understanding, in: Proceedings of the European conference on computer vision (ECCV), 2018, pp. 418--434.

\bibitem{sfnet}
X.~Li, A.~You, Z.~Zhu, H.~Zhao, M.~Yang, K.~Yang, S.~Tan, Y.~Tong, Semantic flow for fast and accurate scene parsing, in: Computer Vision--ECCV 2020: 16th European Conference, Glasgow, UK, August 23--28, 2020, Proceedings, Part I 16, Springer, 2020, pp. 775--793.

\bibitem{detr}
N.~Carion, F.~Massa, G.~Synnaeve, N.~Usunier, A.~Kirillov, S.~Zagoruyko, End-to-end object detection with transformers, in: European conference on computer vision, Springer, 2020, pp. 213--229.

\bibitem{cpvt}
X.~Chu, Z.~Tian, B.~Zhang, X.~Wang, C.~Shen, \href{https://openreview.net/forum?id=3KWnuT-R1bh}{Conditional positional encodings for vision transformers}, in: ICLR 2023, 2023.
\newline\urlprefix\url{https://openreview.net/forum?id=3KWnuT-R1bh}

\bibitem{ghostnet}
K.~Han, Y.~Wang, Q.~Tian, J.~Guo, C.~Xu, C.~Xu, Ghostnet: More features from cheap operations, in: Proceedings of the IEEE/CVF conference on computer vision and pattern recognition, 2020, pp. 1580--1589.

\bibitem{ghostnetv2}
Y.~Tang, K.~Han, J.~Guo, C.~Xu, C.~Xu, Y.~Wang, Ghostnetv2: Enhance cheap operation with long-range attention, Advances in Neural Information Processing Systems 35 (2022) 9969--9982.

\bibitem{unet++}
Z.~Zhou, M.~M. Rahman~Siddiquee, N.~Tajbakhsh, J.~Liang, Unet++: A nested u-net architecture for medical image segmentation, in: Deep Learning in Medical Image Analysis and Multimodal Learning for Clinical Decision Support: 4th International Workshop, DLMIA 2018, and 8th International Workshop, ML-CDS 2018, Held in Conjunction with MICCAI 2018, Granada, Spain, September 20, 2018, Proceedings 4, Springer, 2018, pp. 3--11.

\bibitem{attentionunet}
O.~Oktay, J.~Schlemper, L.~L. Folgoc, M.~Lee, M.~Heinrich, K.~Misawa, K.~Mori, S.~McDonagh, N.~Y. Hammerla, B.~Kainz, et~al., Attention u-net: Learning where to look for the pancreas, arXiv preprint arXiv:1804.03999 (2018).

\bibitem{bihrnet}
Z.~Wu, J.~Zhang, L.~Zhang, X.~Liu, H.~Qiao, Bi-hrnet: A road extraction framework from satellite imagery based on node heatmap and bidirectional connectivity, Remote Sensing 14~(7) (2022) 1732.

\bibitem{zhou2023lithological}
G.~Zhou, W.~Chen, X.~Qin, J.~Li, L.~Wang, Lithological unit classification based on geological knowledge-guided deep learning framework for optical stereo mapping satellite imagery, IEEE Transactions on Geoscience and Remote Sensing (2023).

\bibitem{lei2018ore}
X.-F. Lei, D.-F. Duan, S.-Y. Jiang, S.-F. Xiong, Ore-forming fluids and isotopic (hocs-pb) characteristics of the fujiashan-longjiaoshan skarn w-cu-(mo) deposit in the edong district of hubei province, china, Ore Geology Reviews 102 (2018) 386--405.

\bibitem{xie2015geochemical}
G.~Xie, J.~Mao, Q.~Zhu, L.~Yao, Y.~Li, W.~Li, H.~Zhao, Geochemical constraints on cu--fe and fe skarn deposits in the edong district, middle--lower yangtze river metallogenic belt, china, Ore Geology Reviews 64 (2015) 425--444.

\end{thebibliography}



\end{document}